\documentclass{aa}
    \newcommand*{\scinot}[2]{#1\times10^{#2}}
    \newcommand*{\rd}[2]{\frac{\mathrm{d}#1}{\mathrm{d}#2}}
    
    \newcommand*{\rdil}[2]{\mathrm{d}#1 / \mathrm{d}#2}

    \newcommand*{\bm}[1]{\boldsymbol{\mathbf{#1}}}
    \newcommand*{\uv}[1]{\hat{\bm{#1}}}

    \newcommand*{\p}[1]{\left(#1\right)}
    \newcommand*{\s}[1]{\left[#1\right]}

\usepackage{graphicx, xcolor}
\usepackage{txfonts}
\begin{document}

   \title{The Coupled Tidal Evolution of the Moons and Spins of Warm Exoplanets}

   \author{Yubo Su
          \inst{1,2}
          \and
          Melaine Saillenfest
          \inst{3}
          }

    \institute{
    Department of Astrophysical Sciences, Princeton University, 4 Ivy Lane,
    Princeton, NJ 08540, USA
    \and Canadian Institute for Theoretical Astrophysics, 60 St. George Street,
    Toronto, ON M5S 3H8, Canada\\
    \email{yubo.su@utoronto.ca}
    \and
    LTE, Observatoire de Paris, Universit{\'e} PSL, Sorbonne Universit{\'e},
    Universit{\'e} de Lille, LNE, CNRS, 61 Avenue de l'Observatoire, 75014
    Paris, France\\
    \email{melaine.saillenfest@obspm.fr}
    }

   \date{Received September 15, 1996; accepted March 16, 1997}


  \abstract
  {The Solar System giant planets harbour a wide variety of moons. Among them, the largest moons have moon-to-planet mass ratios of the order of $10^{-4}$. Moons around exoplanets are plausibly similarly abundant, even though most of them are likely too small to be easily detectable with modern instruments. Moons are known to affect the long-term dynamics of the spin of their host planets; however, their influence on warm exoplanets (i.e.\ with moderately short periods of about $10$ to $200$~days), which undergo significant star-planet tidal dissipation, is still unclear.}
  {Here, we study the coupled dynamical evolution of exomoons and the spin dynamics of their host planets, focusing on warm exoplanets.}
  {Analytical criteria give the relevant dynamical regimes at play as a function of the system's parameters. Possible evolution tracks mostly depend on the hierarchy of timescales between the star-planet and the moon-planet tidal dissipations. We illustrate the variety of possible trajectories using self-consistent numerical simulations.}
  {We find two principal results: i) Due to star-planet tidal dissipation, a substantial fraction of warm exoplanets naturally evolve through a phase of instability for the moon's orbit (the `Laplace plane' instability). Many warm exoplanets may have lost their moon(s) through this process. ii) Surviving moons slowly migrate inwards due to the moon-planet tidal dissipation until they are disrupted below the Roche limit. During their last migration stage, moons---even small ones---eject planets from their tidal spin equilibrium. Planets can then converge back to this equilibrium or adopt a new one with a low or high obliquity. Additionally, before their disruption, massive exomoons (with moon-to-planet mass ratios of the order of $10^{-2}$), can maintain their planet in a long-lived high-obliquity state.}
  {The loss of moons through the Laplace plane instability may contribute to disfavour the detection of moons around close-in exoplanets. Moreover, moons (even those that have been lost) play a critical role in the final obliquities of warm exoplanets. Hence, the existence of exomoons poses a serious challenge in predicting the present-day obliquities of observed exoplanets.}


   \maketitle
%

\section{Introduction}

The obliquity of a planet, the angle between its spin and orbital axes, is affected by a number of processes.
In the Solar System, the planets' obliquities range from nearly zero for Mercury to about $180^\circ$ for Venus, including the puzzling $98^\circ$ value of Uranus. This large variety of obliquities is likely the result of distinct combinations of dissipative processes, giant impacts,
as well as spin-orbit and precession resonances \citep[e.g.][]{goldreichpeale1966, safranov1966_uranus,
gold1969_venus,Tremaine_1991,Laskar-Robutel_1993, touma1993chaotic, correia2001_venus,
ward2004I, hamilton2004_2, Ward-Canup_2006, boue2010_uranus,
vokrouhlicky2015tilting,Rogoszinski-Hamilton_2020,Rogoszinski-Hamilton_2021}.

Recently, the first constraints on the obliquities of exoplanets have
become possible \citep{bryan2020obliquity, bryan2021obliquity,
palmabifani2023_obl, poon2024_obl}.
Such measurements have become significantly more feasible with the launch of the
James Webb Space Telescope, which can help infer a planet's spin state
via either asymmetries in its transit light curve \citep[for giant
planets;][]{lammers2024_superpuff, liu2024_kepler167e, lu2024_hip} or direct
measurement of its thermal phase curve
\citep[for close-in planets;][]{greene2023_trappist, hu2024_55cnce,
hammond2024_wasp43b}. The obliquities of exoplanets are important because they are strong markers of the
formation processes and subsequent dynamical evolution of planets, as witnessed by the Solar System planets \citep{levrard2007, fabrycky2007,
Millholland-Laughlin_2018, millholland2019obliquity, millholland2020formation,
su2021dynamics}. Moreover, planetary obliquities play a decisive role in the dynamics -- and possible detection -- of moons and rings around exoplanets (see \citealp{Ohta-etal_2009,tremaine2009satellite}). Because of the close connection between the obliquity and climate of a planet, exoplanetary obliquities are also of critical importance for habitability studies
\citep{kite2011_tidallockclimate, ferreira2014, pennvallis2017, lobo2020, Lobo-Bordoni_2020,
guerrero2024}.

Recent works suggest that the orbital migration of the moons of Jupiter, Saturn, and possibly Uranus are a key ingredient needed to explain the obliquities of these planets \citep{Saillenfest-etal_2020, Saillenfest-etal_2021,Saillenfest-etal_2021b, Wisdom-etal_2022, Dbouk-Wisdom_2023},
and Earth's relatively much
more massive moon also greatly affects its spin evolution (see e.g. \citealp{goldreich1966_moon, laskar1993_moon}).
This is in spite of the comparatively small masses of the moons, which tend to have moon-to-planet mass ratios $\lesssim 10^{-4}$.
The dynamical effect that makes this possible is a secular spin-orbit resonance, a
commensurability between the precession of the planet's spin axis about
its orbit normal (modulated by the moon) and a harmonic of the precession of the planet's
orbit about the planetary system's invariable plane \citep[e.g.][]{colombo1966,
Peale_1969, boue2006precession, Saillenfest-etal_2019,
Saillenfest-etal_2021}.
While exomoons with such small masses are likely undetectable with present-day
instruments, the ongoing interest in exomoon detection
\citep[e.g.][]{kipping2009_moon, teachey2018moon, kipping2022,
yahalomi2024_moon} as well as their aforementioned significant effect on planetary obliquities
motivate careful study of their dynamical impact.

While the general problem of predicting exoplanetary obliquities requires
understanding both planet formation and subsequent dynamical evolution
\citep[e.g.][]{dones1993, millholland_disk, su2020_disk, li2021giant},
and it generally yields a wide range of possible final states, the obliquities of warm exoplanets are in theory able to be much better forecast.
Indeed, the star-planet tidal dissipation generally drives the planet's obliquity to
zero in the absence of secular spin-orbit resonances, or to
one specific state among a few stable equilibria in the presence of resonances
\citep{ward1975tidal, levrard2007, fabrycky2007, Saillenfest-etal_2019,
su2021dynamics, Su-Lai_2022}. The precise outcome of resonance encounter and capture can be modelled as a probabilistic event and be efficiently quantified \citep{henrard1982, Henrard-Murigande_1987, henrard1993, su2021dynamics}.
However, many exoplanets probably have (or have had) moons that continuously migrate as a result of tidal dissipation. The effect of exomoons on previous predictions of the obliquity states of warm planets has not been previously considered.

In this work, we aim to determine the possible spin history of planets subject to: (i)
orbital perturbations from other planets, (ii) a
dissipative torque from its host star, and (iii) an
additional torque from a moon that may migrate
inwards or outwards due to moon-planet tidal dissipation.
These basic ingredients are generic for close-in planets and expected to be very common in
exoplanetary systems.
According to the actual hierarchy of timescales between each dynamical effect, a rich variety of evolutions can be produced, and we investigate a few of interest here.
Note that we restrict our attention to the constant time lag model \citep{alexander1973weak, hut1981tidal} of tidal dissipation: the internal structures of exoplanets are poorly known, and even less is known about their tidal dissipation properties.
Therefore, in this first attempt to explore the coupled dynamics of close-in planets and their exomoons, we will not try to cover the wealth of possible evolution pathways that could be produced by different tidal models, particular rheologies, and dynamical interiors (see e.g. \citealp{FerrazMello_2013, Boue-etal_2016, fuller2016resonance, Ragazzo-Ruiz_2017, FerrazMello-etal_2020, Boue_2020}).

In Sect.~\ref{sec:theory}, we recall the dynamical mechanisms at play and outline the basic equations that govern the spin, orbit, lunar, and tidal evolutions.
We then describe trajectories obtained in three different regimes according to the ratio between the timescale of star-planet tidal dissipation $\tau_{\rm
p}$ and that of moon-planet tidal dissipation $\tau_{\rm m}$.
In Sect.~\ref{sec:longlived} we first study the regime where $\tau_{\rm m}\lesssim \tau_{\rm p}$ and find that the planet spindown enhances obliquity excitation due to lunar migration. This section connects our work to previous studies which have focussed on the regime $\tau_{\rm m}\ll\tau_{\rm p}$.
In Sect.~\ref{sec:e1_instab}, we study the early phases of the regime where $\tau_{\rm m} \gg \omega_{\rm p}$ and show that moons can be lost more frequently than previously thought.
In Sect.~\ref{sec:despun}, we study the late phases of the same regime $\tau_{\rm m} \gg \omega_{\rm p}$, where the eventual runaway tidal inspiral of the moon breaks the planet out of its tidal equilibrium state.
We summarize and discuss our results in Sect.~\ref{sec:conclusion}.

\section{Theoretical background}\label{sec:theory}

In this section, we will describe some simple, and sometimes approximate, analytical results that are useful towards understanding the combined spin and orbital evolution of planets and their moons.
Later in this paper, we will use these results to analyze numerical simulations that are not subject to the same approximations.

\subsection{Spin-orbit Precession and Secular Resonances}\label{ssec:res}

In this section, we follow the formalism of \citet{Saillenfest-etal_2019} and \citet{su2021dynamics}, with some changes in notation.
We consider a star hosting several planets that perturb each other's orbits.
We study the spin dynamics of one of these planets denoted with the subscript ``$\mathrm{p}$'' (to distinguish quantities related to the planet from quantities related to its moon, see below).
We assume that the planetary system is long-term stable, then the orbital inclination of the planet can be written (at least locally in time) as a convergent quasi-periodic series truncated to $N$ terms:
\begin{equation}\label{eq:zeta}
   \zeta_\mathrm{p} \equiv \sin\frac{I_\mathrm{p}}{2}\exp(i\,\Omega_\mathrm{p}) = \sum_{j=1}^N\Upsilon_j\exp[i\,\phi_j(t)]\,.
\end{equation}
Here, $I_\mathrm{p}$ is the orbital inclination of the planet and
$\Omega_\mathrm{p}$ is its longitude of ascending node measured in some inertial
reference frame.
$\Upsilon_j$ is a real positive constant and $\phi_j(t)=\nu_j\,t + \phi_j^{(0)}$
evolves linearly with time $t$ with frequency $\nu_j$.
Such a quasi-periodic decomposition can be obtained from an analytical theory
(e.g.\ the Lagrange-Laplace system, see \citealp{Murray-Dermott_1999}) or from a
Fourier analysis of a numerical solution (see e.g.\
\citealp{Laskar_1988,Laskar_1990}).
Equation~\eqref{eq:zeta} defines the motion of the unitary orbital angular momentum
vector $\uv{l}_\mathrm{p}$ of the planet due to interactions with the other planets in the system via
\begin{equation}
    \uv{l}_\mathrm{p}=(\sin\Omega_{\rm p}\sin I_{\rm p},-\cos\Omega_{\rm p}\sin
I_{\rm p},\cos I_{\rm p})^\mathrm{T}.
\end{equation}

Next, we study the evolution of the unitary spin angular momentum of the planet, denoted $\uv{s}_\mathrm{p}$.
In the absence of moons, the dominant torque on $\uv{s}_\mathrm{p}$ is due to
the attraction of the star on the planet's equatorial bulge.
At quadrupolar order, and upon averaging over the fast orbital and spin angles,
the equation of motion for $\uv{s}_\mathrm{p}$ is
\begin{equation}\label{eq:dsdt}
   \frac{\mathrm{d}\uv{s}_\mathrm{p}}{\mathrm{d}t} = \alpha\left(\uv{s}_\mathrm{p}\cdot\uv{l}_\mathrm{p}\right)\left(\uv{s}_\mathrm{p}\times\uv{l}_\mathrm{p}\right)\,,
\end{equation}
where
\begin{equation}\label{eq:alpha}
   \alpha = \frac{3}{2}\frac{\mathcal{G}m_\star}{a_\mathrm{p}^3(1 - e_\mathrm{p}^2)^{3/2}}\frac{J_2}{\omega\lambda}\,,
\end{equation}
(see e.g. \citealp{Laskar-Robutel_1993}, Eq. 2). Here, $\mathcal{G}$ is the gravitational constant, $m_\star$ is the mass of the
star, $a_\mathrm{p}$ and $e_\mathrm{p}$ are the semi-major axis and eccentricity
of the planet's orbit around the star, $J_2$ is the second zonal
gravity coefficient of the planet, $\omega$ is its spin rate, and $\lambda$ is
its normalised polar moment of inertia. The quantities $J_2$ and $\lambda$ are
defined using the same normalising radius
$R_\mathrm{p}$ of the planet (e.g.\ its equatorial radius at time $t=0$).

In Eq.~\eqref{eq:dsdt}, the spin-axis motion of the planet is forced by
its orbital motion through the explicit time dependence of $\uv{l}_\mathrm{p}$
coming from Eq.~\eqref{eq:zeta}.
As the orbital angular momentum of the planet is much larger that its spin
angular momentum, the back reaction of the planet's spin-axis motion on the planet's
orbit is negligible.

The obliquity $\theta$ of the planet is defined as $\cos\theta =
\uv{s}_\mathrm{p}\cdot\uv{l}_\mathrm{p}$.
Because of the motion of the planet's orbital plane, the planet's spin-axis dynamics is
affected by secular spin-orbit resonances, that is, by resonances between the
precession of the spin axis and one or several harmonics appearing in
Eq.~\eqref{eq:zeta}.
As detailed by \citet{Saillenfest-etal_2019}, the strongest resonances are of
order $1$ in the amplitudes $\Upsilon_j$.
If the planet is trapped in such a resonance with harmonic $j=k$, then the
resonance angle $\sigma = \psi + \phi_k$ oscillates around a fixed value, where
$\psi$ is the precession angle of the planet's spin axis \citep[following the
convention of][]{Laskar-Robutel_1993, Saillenfest-etal_2019}.
In other words, the planet's obliquity $\theta$ evolves such that the relation
$\alpha\cos\theta + \nu_k \approx 0$ is maintained.
This kind of equilibrium is called a `Cassini State' (see e.g.\
\citealp{Peale_1969,Henrard-Murigande_1987}).
Depending on the value of parameters, there can be up to four distinct Cassini
states.

Neglecting terms of order $\Upsilon_j^4$ and beyond, we introduce the non-dimensional variables $\gamma = f_\gamma/\alpha$ and $\beta= f_\beta/\alpha$, where
\begin{equation}
   \begin{aligned}
      f_{\gamma} &= -\left(\nu_k - 2\sum_{j=1}^N\nu_j\Upsilon_j^2\right)\,, \\
      f_{\beta} &= -\Upsilon_k\left(2\nu_k + \nu_k\Upsilon_k^2 - 2\sum_{j=1}^N\nu_j\Upsilon_j^2\right)\,.
   \end{aligned}
\end{equation}
As an immediate property of the Lagrange-Laplace system, we expect dominant terms in Eq.~\eqref{eq:zeta} to have negative frequencies $\nu_j$ if the planetary system is roughly coplanar (see e.g. \citealp{Laskar_1990}). This means that for the strongest resonances, $\gamma$ and $\beta$ are both positive.

At the vicinity of resonance with harmonic $k$, Eq.~\eqref{eq:dsdt} can be rewritten component-wise as (see \citealp{Saillenfest-etal_2019}, Eq. 16)\footnote{We note that the sign of $\sigma$ here is opposite to that of \citet{su2021dynamics}, so
our equations of motion are consistent with theirs.}
\begin{align}
    \left.\rd{\theta}{t}\right|_\mathrm{res} &= -\alpha\beta\sin\sigma\,,\label{eq:dthetares}\\
    \left.\rd{\sigma}{t}\right|_\mathrm{res} &= \alpha\cos\theta - \alpha\gamma - \alpha\beta\cot\theta \cos\sigma\,.\label{eq:dsigmares}
\end{align}
The Cassini states are the equilibrium points of this system of two equations.
For $\gamma^{2/3} + \beta^{2/3} \geqslant 1$, there are only two distinct
Cassini states and no resonance can be defined (i.e.\ there is no separatrix).
For $\gamma^{2/3} + \beta^{2/3} < 1$, there are four Cassini states, and a
separatrix encloses the resonance region.
The equilibrium points corresponding to the resonance centre is called `Cassini
state 2'.

If the orbital inclination dynamics of the planet is dominated by one single
harmonic in Eq.~\eqref{eq:zeta} with frequency $\nu$ and amplitude
$\Upsilon=\sin(I_\mathrm{p}/2)$, then, only one secular spin-orbit resonance exists.
For this isolated resonance, $\gamma$ and $\beta$ are simplified to $\gamma =
-\nu\cos I_\mathrm{p}/\alpha$ and $\beta = -\nu\sin I_\mathrm{p}/\alpha$, and we
recover the formulas given by \citet{su2021dynamics}.
In general, however, the orbital dynamics of planets are composed of many
harmonics that create a web of resonances.
These resonances can either be isolated (see e.g.\
\citealp{Saillenfest-etal_2020}) or overlap and create chaotic zones (see
\citealp{Laskar-Robutel_1993}).

\subsection{Tidal Dissipation and Stable Equilibria}\label{ssec:dissip}

We consider that the planet is relatively close to its host star, with an
orbital period smaller than a few hundred days.
In this case, the substantial star-planet tidal interactions produce redistributions
of mass inside the planet, leading to a dissipation of energy.
Using the standard `constant time lag' weak friction theory of the equilibrium
tide \citep{alexander1973weak, hut1981tidal, lai2012}, the spin rate $\omega$
and obliquity $\theta$ of the planet evolve due to tidal dissipation, following
\begin{align}
    \left.\rd{\omega}{t}\right|_\mathrm{tides}
        &= \frac{\omega}{\tau_{\rm p}} \s{\frac{2n_{\rm p}}{\omega}\cos\theta - \left(1 + \cos^2\theta\right)}\,,\label{eq:domegatides}\\
    \left.\rd{\theta}{t}\right|_\mathrm{tides} &=
        -\frac{1}{\tau_{\rm p}}\sin \theta\p{2\frac{n_\mathrm{p}}{\omega} - \cos\theta}\,,\label{eq:dthetatides}
\end{align}
where $n_\mathrm{p}$ is the mean motion of the planet, and
$\tau_\mathrm{p}$ is the tidal timescale.
Assuming that the planet behaves as a self-gravitating fluid, $\tau_\mathrm{p}$
is given by
\begin{align}
    \frac{1}{\tau_{\rm p}} ={}& \frac{3k_2n_{\rm p}\Delta t}{2\lambda}
        \p{\frac{m_\star}{m_{\rm p}}}
        \p{\frac{R_{\rm p}}{a_{\rm p}}}^3 n_{\rm p}\,,\label{eq:taup}\\
        ={}& \frac{1}{1\; \mathrm{Gyr}}
            \p{\frac{k_2n_{\rm p}\Delta t/\lambda}{10^{-4}}}
            \p{\frac{M_\star}{M_{\odot}}}^{3/2}
            \p{\frac{m_{\rm p}}{10M_{\oplus}}}^{-1}\nonumber\\
        &\times \p{\frac{R_{\rm p}}{2R_{\oplus}}}^3
            \p{\frac{a_{\rm p}}{0.4 \mathrm{au}}}^{-9/2}.\nonumber
\end{align}
where $\Delta t$ is
the tidal time lag\footnote{We note that the tidal lag time is sometimes associated with an ``effective'' tidal quality factor $Q$ via $2n_{\rm p}\Delta t \sim 1/Q$ \citep{lai2012, lu2023rebound}.} and $k_2$ is the second tidal Love number of the planet.
Due to planetesimal accretion for small planets \citep{dones1993, miguel2010_impacts}, and to a combination of runaway accretion and magnetic despinning for giant planets \citep{takata1996_giantrotation, batygin2018_giantrotation}, we can expect planets to be born with a spin rate up to critical rotation. Therefore, their initial spin
rates $\omega_0$ can be substantially supersynchronous:
\begin{align}
    \frac{\omega_0}{n_{\rm p}}
        \lesssim \frac{\omega_{\rm crit}}{n_{\rm p}}
            &= \sqrt{\frac{m_{\rm p}a_{\rm p}^3}{M_\star R_{\rm p}^3}}
            \nonumber\\
        &=
            1800
            \p{\frac{(m_{\rm p}/m_\star)}{(10 M_\oplus/M_\odot)}}^{1/2}
            \p{\frac{(a_{\rm p}/R_{\rm p})}
                {(0.4\;\mathrm{au} / 2R_\oplus)}}^{3/2}\,,
                \label{eq:omega_p0}
\end{align}
where $\omega_\mathrm{crit} = \sqrt{\mathcal{G}m_\mathrm{p}/R_\mathrm{p}^3}$ is
the critical spin rate of the planet above which its centrifugal potential
overwhelms its gravitational potential. As a consequence, the time it takes for a planet to become tidally synchronized (i.e.\ to reach $\omega\approx n_\mathrm{p}$) is typically
several $\tau_{\rm p}$.

As $\omega$ evolves due to tidal dissipation, the planet's oblateness evolves following its hydrostatic equilibrium figure \citep{eggleton2006book},
\begin{equation}\label{eq:J2}
   J_2 = \frac{k_2}{3}\left(\frac{\omega}{\omega_\mathrm{crit}}\right)^2\,,
\end{equation}
where $k_2$ is the hydrostatic Love number
of the planet.

The variations in $\omega$ mean that the coefficient $\alpha$ in
Eq.~\eqref{eq:alpha} is not constant; yet, the change in $\alpha$ due to tidal
dissipation is considered to be slow enough so that Eqs.~\eqref{eq:dthetares}
and~\eqref{eq:dsigmares} still hold, with $\alpha$ treated as a slowly-varying
parameter.
Hence, the complete set of equations that governs the spin-axis
dynamics of the planet due to secular spin-orbit resonances and star-planet
tidal dissipation is obtained by adding both contributions (see
Eq.~\ref{eq:dthetares}-\ref{eq:dsigmares}
and~\ref{eq:domegatides}-\ref{eq:dthetatides}), resulting in a system of three
equations for the variables $(\omega,\theta,\sigma)$.

The equilibria of this system are the `tidal Cassini equilibria' (or tCE\@; see
\citealp{su2021dynamics}).
In the limit of weak dissipation, the locations of tCE only depend on
$\gamma_\mathrm{sync} = f_\gamma/\alpha_\mathrm{sync}$ and $\beta_\mathrm{sync} =
f_\beta / \alpha_\mathrm{sync}$, where $\alpha_\mathrm{sync}$ is the value of
$\alpha$ computed at $\omega = n_\mathrm{p}$.
For a configuration of the system to be a tCE, Eq.~\eqref{eq:domegatides}
implies that the planet must have reached the pseudo-synchronous spin state,
that is,
\begin{equation}\label{eq:omegapseudo}
   \left.\frac{\omega}{n_\mathrm{p}}\right|_{\dot\omega=0} =
       \frac{2\cos\theta}{1 + \cos^2\theta}\,.
\end{equation}
This condition requires the planet's spin at equilibrium to be prograde, that
is, $\theta\leqslant90^\circ$.
For $\gamma_\mathrm{sync}^{2/3} + \beta_\mathrm{sync}^{2/3} \geqslant 1$, there
is only one tCE, called $\mathrm{tCE}_2$, which corresponds to the Cassini state~2 at
pseudo-synchronisation\footnote{
Strictly speaking, the spin equilibrium has a small nonzero $\sigma$, and hence could be called a \emph{modified} Cassini state to distinguish from the non-dissipative equilibrium \citep{su2021dynamics}.
Since we study only the dissipative dynamics, we refer to them as just Cassini states for the remainder of the work.}.
For $\gamma_\mathrm{sync}^{2/3} + \beta_\mathrm{sync}^{2/3} < 1$, there are two
tCE, called $\mathrm{tCE}_1$ and $\mathrm{tCE}_2$, which correspond to the Cassini states 1 and 2 at
pseudo-synchronisation.
The one with higher obliquity is $\mathrm{tCE}_2$; it is located at the centre of the
secular spin-orbit resonance.
In order for the Cassini states 1 and 2 to exist at $\theta\leqslant 90^\circ$,
the frequency $\nu_k$ of the resonance needs to be negative, which implies that
$\gamma$ and $\beta$ are positive.

In the limit $\alpha \gg |\nu_k|$, that is, $\gamma\ll 1$ and $\beta\ll 1$, the
Cassini state 2 is located at $\cos\theta \approx \gamma$. By virtue
of~\eqref{eq:omegapseudo}, the obliquity and spin rate at $\mathrm{tCE}_2$ are therefore
\begin{align}
   \cos\theta_{\rm tCE2} &\approx \sqrt{\frac{\gamma_\mathrm{sync}}{2}}\,,
   \label{eq:def_tce2_qapprox}\\
   \frac{\omega_\mathrm{tCE2}}{n_\mathrm{p}} &\approx \sqrt{2\gamma_\mathrm{sync}}\,.
   \label{eq:def_tce2_approx}
\end{align}

\subsection{Tidal Resonance Breaking}\label{ssec:tce_break}

For strong enough dissipation (i.e.\ small enough tidal timescale $\tau_{\rm
p}$), the high-obliquity Cassini state 2 is no longer stable to tidal dissipation \citep[specifically, it undergoes a saddle node bifurcation with the unstable Cassini State 4][]{levrard2007, fabrycky2007, su2021dynamics}.
Indeed, Eqs.~\eqref{eq:dthetares} and~\eqref{eq:dthetatides} imply that a
necessary condition for $\rdil{\theta}{t}$ to have a
zero is
\begin{equation}
    f_{\beta}\geq \frac{1}{\tau_{\rm p}}
        \sin\theta\p{2\frac{n_{\rm p}}{\omega} - \cos \theta}.
\end{equation}
Evaluating this equation for $\omega$ and $\theta$ given by
Eqs.~\eqref{eq:def_tce2_qapprox} and~\eqref{eq:def_tce2_approx}, we obtain that
\begin{equation}
   \sqrt{\frac{\gamma_\mathrm{sync}}{2}} \gtrsim \frac{1}{f_{\beta}\tau_{\rm p}}\,.\label{eq:tCE2_break}
\end{equation}
This is a necessary condition for $\mathrm{tCE}_2$ to be stable to tidal dissipation.
As such, if the tidal timescale $\tau_{\rm p}$ becomes sufficiently short, or if
$\gamma_\mathrm{sync}$ becomes sufficiently small, then the planet is ejected
out of the high-obliquity $\mathrm{tCE}_2$, whereupon it settles into the low-obliquity $\mathrm{tCE}_1$.

\subsection{Precession Dynamics with a Moon: General Theory}\label{ssec:moon}

We next consider the spin dynamics of the planet as affected by a moon in orbit around it. Since the moon's eccentricity is expected to
damp relatively rapidly (e.g.\ \citealp{goldreich1963eccentricity}), we consider for now that the
moon's orbit is circular. Otherwise, call the moon's mass $m_{\rm m}$, its
semimajor axis $a_{\rm m}$, and the inclination of its orbit relative to the
planet's equatorial plane $i_{\rm m}$.

\begin{figure*}
    \centering
    \includegraphics[width=\linewidth]{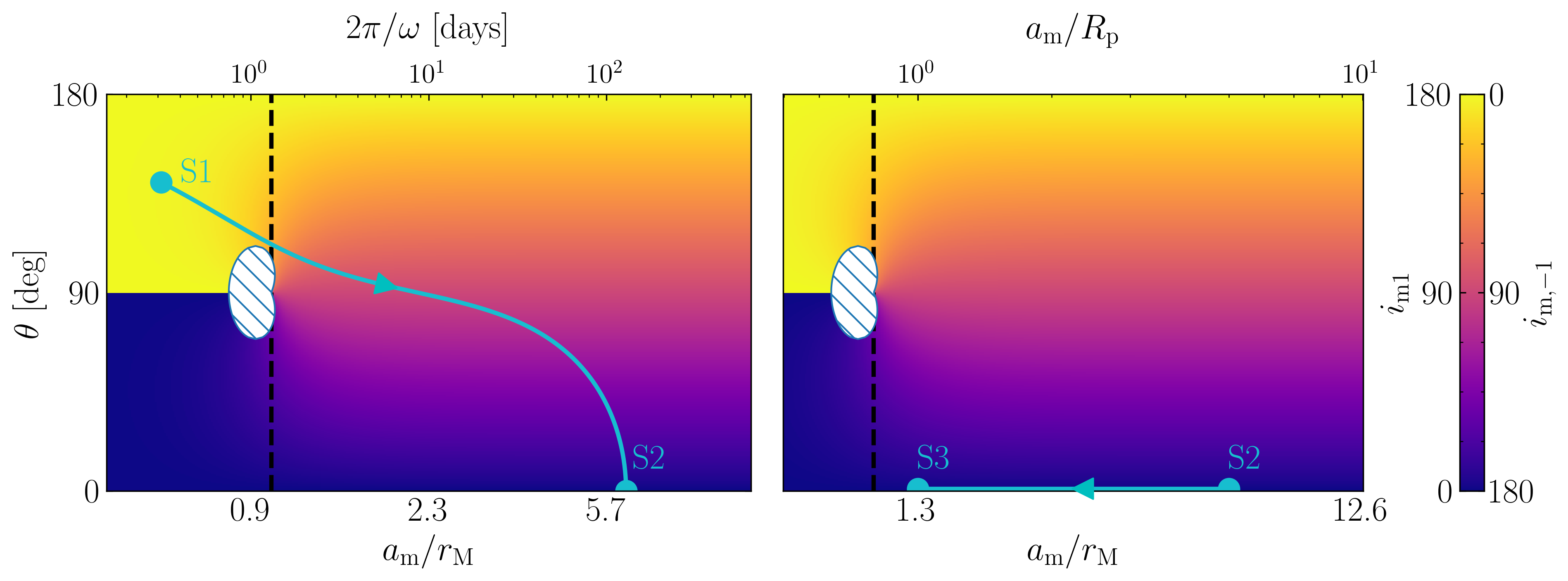}
    \caption{
    Orbital inclinations of Laplace state $\mathrm{P}_1$ and its twin $\mathrm{P}_{-1}$ (colorbar, given by Eq.~\ref{eq:i_p1},~\ref{eq:def_im1prime}) as functions of $a_{\rm m} / r_{\rm M}$ and the planet's obliquity $\theta$.
    In both panels, the location of
    $a_{\rm m} = r_{\rm M}$ is denoted with a vertical black dashed line, and the $\mathrm{E}_1$ region (where $\mathrm{P}_1$ and $\mathrm{P}_{-1}$ are unstable) is shown by a white hatched zone.
    Across the two panels, a representative evolutionary history is shown via
    the three cyan points and arrows (S1, S2, S3).
    While the inclinations of $\mathrm{P}_1$ and $\mathrm{P}_{-1}$ only depend on $\theta$ and $a_{\rm m} / r_{\rm M}$, we also illustrate the evolution using the following concrete parameters for interpretability: $a_{\rm p} =
    0.4\;\mathrm{au}$, $e_{\rm p}=0$, $R_{\rm p} = 2R_\oplus$, and $m_{\rm p} = 10M_\oplus$;
    We then denote the corresponding values of the planet's spin rate along the top-left axis (relevant to
    planetary tidal evolution) and the lunar semi-major axis along the top-right axis (relevant
    to lunar migration).
    In the left panel, the planet is initially rapidly rotating with a period of
    $7.5$ hours, and the moon is located at $a_{\rm m} = 5R_{\rm p}$ (S1).
    Then, as the planet despins and the obliquity damps, $r_{\rm M}$ decreases
    while the moon is still located at $a_{\rm m} = 5R_{\rm p}$, and the system reaches S2 (we have neglected lunar tidal migration for illustrative purposes).
    Going from S1 to S2, the inclination of the lunar orbit smoothly flips from prograde to
    retrograde (assuming that the moon occupies $\mathrm{P}_{-1}$).
    In the right panel, as the moon inspirals, starting from S2, $a_{\rm m}$
    decreases until the moon is disrupted by the planet when $a_{\rm m} \simeq R_{\rm p}
    \gtrsim r_{\rm M}$ (S3).
    }\label{fig:laplace_p1}
\end{figure*}

The orbit of a moon is affected by secular precessional torques from both the
planet's equatorial bulge and the distant host star.
In the limit of a massless moon, there are six equilibria of these precession equations, known
as the Laplace states \citep{tremaine2009satellite, saillenfest2021future}. We note them $\mathrm{P}_1$, $\mathrm{P}_2$, $\mathrm{P}_3$, and $\mathrm{P}_{-1}$, $\mathrm{P}_{-2}$, $\mathrm{P}_{-3}$, where the negative indexes denote the twin equilibria under the transformation
$\uv{l}_{\rm m} \to -\uv{l}_{\rm m}$, where $\uv{l}_{\rm m}$ is the moon's orbit
normal.
Since regular moons probably form in an environment where damping is efficient due to gas drag and/or mutual collisions, their initial orbital plane is thought to lie very close to a stable equilibrium near the equator of their host planets.
The Laplace states corresponding to such an equilibrium are $\mathrm{P}_1$ and $\mathrm{P}_{-1}$, which are denoted the ``circular
coplanar Laplace
equilibrium'' by \citet{tremaine2009satellite}. In the limit of a massless moon, the inclination of $\mathrm{P}_1$ is given by
(\citealp{saillenfest2021future}; see \citealp{boue2006precession} for the
general expression)
\begin{equation}
    i_{\rm m1} = \frac{1}{2}
        \s{\pi + \mathrm{atan2}\p{-\sin 2\theta,
            -(r_{\rm M} / a_{\rm m})^5 - \cos\p{2\theta}}},\label{eq:i_p1}
\end{equation}
where
\begin{equation}
    r_{\rm M}^5 \equiv 2\frac{m_{\rm p}}{m_\star}
        J_2 R_{\rm p}^2a_{\rm p}^3(1-e_\mathrm{p}^2)^{3/2},
   \label{eq:rM}
\end{equation}
is the ``midpoint radius'', related to the canonical Laplace radius $r_{\rm L}$ \citep{tremaine2009satellite}
via $r_{\rm M}^5 = 2r_{\rm L}^5$. We prefer to use $r_{\rm M}$ since it is the
lunar semimajor axis at which $i_{\rm m1}$ lies exactly halfway between the planet's equatorial and orbital planes (i.e. $i_{\rm m1} = \theta / 2$ for $\theta < 90^\circ$).
For clarity, we call $i_{\rm m,-1}$ the inclination of $\mathrm{P}_{-1}$ (i.e. the ``twin'' state to $\mathrm{P}_1$ with reversed lunar orbital motion). Its value is
\begin{equation}
    i_{\rm m,-1} \equiv 180^\circ - i_{\rm m1}.\label{eq:def_im1prime}
\end{equation}
The behavior of $i_{\rm m1}$ as functions of the evolving system parameters is illustrated in Fig.~\ref{fig:laplace_p1}.
Depending on the formation process of regular moons (e.g. in a primordial circumplanetary disk, or as a second generation moon system following a tilting impact; see e.g.\ \citealp{Morbidelli-etal_2012}), their orbits can be initially prograde or retrograde to their planet's spin.
If we consider a prograde moon around a retrograde spinning planet (i.e. $\theta > 90^\circ$, such as the major moons of Uranus), this moon occupies the $\mathrm{P}_{-1}$ state and has an inclination equal to $i_\mathrm{m,-1}$. Following Fig.~\ref{fig:laplace_p1}, we see that such a moon evolves into a
retrograde orbit by the time the planet has despun (left panel in conjunction
with Eq.~\ref{eq:def_im1prime}; see also Fig.~3-4 by \citealp{saillenfest2021future}). This is because the oscillations of the moon's orbital pole around the stable equilibrium occur on a much shorter timescale than the despinning of the planet, such that the moon adiabatically follows the equilibrium.

For a moon located in the $\mathrm{P}_1$ or $\mathrm{P}_{-1}$ state, the
spin-axis precession of the planet is modified by replacing $\alpha$ in Eq.~\eqref{eq:dsdt} by
\begin{equation}
   \alpha' = \frac{3}{2}\frac{\mathcal{G}m_\star}{a_\mathrm{p}^3(1 - e_\mathrm{p}^2)^{3/2}}\frac{J_2'}{\omega\lambda'}\,,
   \label{eq:alphap_full}
\end{equation}
where we define the effective $J_2$ and $\lambda$ as \citep{French-etal_1993,boue2006precession,
saillenfest2021future}
\begin{align}
    J_2' &= J_2 + \frac{1}{2}\frac{m_{\rm m}a_{\rm m}^2}{m_{\rm p}R_{\rm p}^2}
        \frac{\sin\p{2\theta - 2i_{\rm m}}}{\sin\p{2\theta}}
        \label{eq:J2p},\\
    \lambda' &= \lambda + \frac{m_{\rm m}a_{\rm m}^2}{m_{\rm p}R_{\rm p}^2}
        \frac{n_{\rm m}}{\omega}
        \frac{\sin\p{\theta - i_{\rm m}}}{\sin\theta}\label{eq:lambdap},
\end{align}
and $i_\mathrm{m}$ is equal to either $i_\mathrm{m1}$ or $i_\mathrm{m,-1}$. Here, $n_{\rm m}$ is the moon's mean motion. We note that this expression differs from that used by \cite{ward2004I, millholland2019obliquity}---see
discussion by \cite{saillenfest2021future}\footnote{Equation~\eqref{eq:lambdap} also slightly differs from Eq.~(39) of
\cite{saillenfest2021future}, as we have removed an absolute value. This absolute value was put by \cite{saillenfest2021future} to enforce the moon to always be prograde to the planet's spin in their calculations, without the need to change the expression of $i_\mathrm{m}$ from $i_\mathrm{m1}$ to $i_\mathrm{m,-1}$. We drop this ambiguous notation here and use the classic expression with parentheses (see e.g. \citealp{French-etal_1993}).}.
For tidally despun planets, we stress that both $\lambda'$ and $J_2'$ generally differ substantially from $\lambda$ and $J_2$.

As first pointed out by \cite{tremaine2009satellite}, the $\mathrm{P}_1$ and $\mathrm{P}_{-1}$ state of the moon are linearly unstable to eccentricity growth in a closed region of the space $(a_\mathrm{m}/r_\mathrm{M},\theta)$. We call this region $\mathrm{E}_1$; it has a cardioid-like shape, whose border has a simple analytical expression (see \citealp{saillenfest2021future} and the hatched region in Fig.~\ref{fig:laplace_p1}). Qualitatively, the $\mathrm{E}_1$ region lies at $a_{\rm m} \sim r_{\rm M}$ for planet's obliquity values $\theta \in [68.9^\circ, 111.1^\circ]$. When smoothly entering $\mathrm{E}_1$ via a slow change of parameters, trajectories on the circular $\mathrm{P}_1$ and $\mathrm{P}_{-1}$ equilibria transition to stable eccentric equilibria, which then become linearly unstable deeper in the $\mathrm{E}_1$ region.

\subsection{Precession Dynamics with a Moon: Synchronously Spinning Planet}\label{ssec:moonsync}

Here, we derive some special cases of the results presented in the previous
section for applications towards planets that have tidally synchronized.
First, for $e_\mathrm{p}=0$ and $m_\mathrm{p}\ll m_\star$, the midpoint radius $r_{\rm M}$ can be approximated by
\begin{equation}
    r_{\rm M}^5
        \approx \frac{2 k_{\rm 2}}{3} \p{\frac{\omega}{n_{\rm p}}}^2 R_{\rm p}^5.
        \label{eq:rM_sync}
\end{equation}
As such, we can see that as planets approach spin-orbit synchronization, $r_{\rm
M} \lesssim R_{\rm p}$, so their moons are always exterior to the midpoint
radius (by merit of being outside of the planet).

Next, we seek expressions for $\alpha'$ in this limit. It follows from
Eqs.~(\ref{eq:i_p1},~\ref{eq:rM}) that when $a_{\rm m} \gg
r_{\rm M}$,
\begin{align}
    \frac{\sin\p{2\theta - 2i_{m}}}{\sin \p{2\theta}}
        &\approx \frac{2k_2}{3}\frac{m_{\rm p}}{m_\star}
        \frac{R_{\rm p}^2a_{\rm p}^3}{a_{\rm m}^5}
            \p{\frac{\omega}{\omega_{\rm crit}}}^2,\\
    \alpha'
        &\approx
            \alpha_0 \s{1 + \frac{m_{\rm m}a_{\rm p}^3}{m_\star a_{\rm m}^3}}\frac{\lambda'}{\lambda}
                \label{eq:alphap_slow}\,.
\end{align}
Since $\lambda \gg J_2$, the relative change from $\lambda$ to $\lambda'$ is often modest, even while the relative change from $J_2$ to $J_2'$ is significant (Eqs.~\ref{eq:J2p} and~\ref{eq:lambdap}).
Accordingly, $\alpha'$ only varies moderately as the planet spins down, and it is constant in the limit $\lambda' \approx \lambda$.

\subsection{Lunar Orbit Evolution}\label{ssec:moontide}

Throughout this paper, we consider cases for which the planet's spin evolution due to the tide raised by the star is either slower or faster than the moon's orbital
evolution due to the tide it raises on the planet. The variety of possible evolution pathways depend on the hierarchy between these two timescales.
By looking at the major moons in the Solar System, we know that regular moons can be found at distances $a_\mathrm{m}$ between a few and a few tens of $R_\mathrm{p}$. The most distant major moons may have reached their current distances through tidal migration (see e.g. \citealp{goldreichsoter1966,lainey2020resonance}).
Adopting the same constant time lag model as in Sect.~\ref{ssec:dissip}, but for the lunar tide, the lunar orbit
evolves following:
\begin{align}
    \frac{1}{a_{\rm m}}\rd{a_{\rm m}}{t}
        ={}&
            -\frac{1}{\tau_{\rm m}}\p{1 - \frac{\omega}{n_{\rm m}}\cos i_{\rm m}},\\
    \frac{1}{\tau_{\rm m}}
        ={}& (6k_2n_{\rm p}\Delta t) \frac{m_{\rm m}}{m_{\rm p}}
            \p{\frac{R_{\rm p}}{a_{\rm m}}}^5
            n_{\rm m},\\
        ={}& \frac{1}{14\;\mathrm{Gyr}}
            \p{\frac{4k_2n_{\rm p}\Delta t}{10^{-4}}}
            \p{\frac{m_{\rm m} / m_{\rm p}}{10^{-4}}}
            \p{\frac{R_{\rm p} / a_{\rm m}}{0.1}}^{13/2}\nonumber\\
        &\times \p{\frac{m_{\rm p}}{10M_\oplus}}^{1/2}
            \p{\frac{R_{\rm p}}{2R_{\oplus}}}^{-3/2}.\label{eq:taum}
\end{align}
The backreaction of the moon's orbital evolution on the planet's spin tends to drive the planet towards synchronization
and alignment with the orbit of its moon on characteristic timescale
\begin{equation}
    \frac{1}{\tau_{\rm p, m}}
        = \frac{3k_2n_{\rm p}\Delta t}{2\lambda}\p{\frac{m_{\rm m}}{m_{\rm p}}}^2
            \p{\frac{R_{\rm p}}{a_{\rm m}}}^3
            n_{\rm m}.\label{eq:tau_pm}
\end{equation}
For the fiducial values considered in Eq.~\eqref{eq:taum}, the lunar tide typically results in evolution on longer timescales than the stellar tide (i.e. $\tau_\mathrm{p,m}>\tau_\mathrm{p}$).
However, for more sophisticated tidal processes, lunar inspiral can be
accelerated compared to the na\"ive model considered here.
For instance, resonance locks are thought to be responsible for the anomalously
rapid migration of the moons of Saturn (\citealp{fuller2016resonance,
lainey2020resonance}; but see also \citealp{Jacobson2022Saturn}).
We note here that we also neglect the effect of \emph{cross tides}, a phenomenon that arises when two bodies raise tidal perturbations on the same primary \citep{goldreich1966_moon, touma1994_ctides, desurgy1997_ctides}.
We justify this omission in Appendix~\ref{app:cross_tides}.

\subsection{Numerical procedure}\label{sec:numcode}

The previous subsections outline the dominant effects at play, allowing us to get an intuition of the possible evolutionary tracks of the planet and its moon. For the case studies analyzed in the following sections, we systematically validate our analytical approximations and explore the dynamics in more details using a self-consistent numerical model.

Our numerical experiments below are produced with the coupled equations of motion of the planet's spin axis and the orbit of its moon. The methodology is described in Appendix~C of \cite{Saillenfest-etal_2022}: It is based on the secular equations of motion expanded to quadrupole order developed by \cite{Correia-etal_2011}, which are valid for arbitrary eccentricities and inclinations for the planet and the moon. The orbital evolution of the moon is integrated self consistently, with no assumption regarding its eccentricity or inclination. In order to reproduce the perturbations from additional planets in the system, the orbital evolution of the planet is made to vary with quasi-periodic series as in Eq.~\eqref{eq:zeta}.

The tidal dissipation due to the stellar torque acting on the planet's spin axis is included using the constant time lag theory (see Sect.~\ref{ssec:dissip} and Eq.~22 of \citealp{Correia-etal_2011}). As the orbital angular momentum of the planet is much larger than its spin angular momentum, we neglect any dissipative effect on the planet's orbit (i.e. the planet does not migrate on its orbit around the star).

Depending on the section, the tidal dissipation due to the moon's torque is included in a different way. In Sect.~\ref{sec:longlived}, the moon's orbital migration (i.e. the evolution of its semi-major axis) is introduced as a pre-defined function of time, and we neglect its effect on the planet's despinning. In Sects.~\ref{sec:e1_instab} and \ref{sec:despun}, the moon's migration and tidal despinning of the planet are included self consistently: the moon migrates as a response of angular momentum transfer
(see Eq.~20 of \citealp{Correia-etal_2011}). Each setting is described in the corresponding section.

As the equations of motion are averaged over the mean anomaly of the moon and the mean anomaly of the planet, non-secular effects such as evection and eviction-like resonances are not included in our numerical experiments (see \citealp{Touma-Wisdom_1998,Vaillant-Correia_2022}). This choice is justified in Appendix~\ref{asec:evec}. We show that evection and eviction-like resonances do not play a role here because the star-planet tidal dissipation despins the planet to near spin-orbit synchronization; As a consequence, these resonances are shifted to distances below the planet's own radius before the moon has a chance to reach them.

Our numerical experiments do not include tidal dissipation within the moon's interior or the evolution of the moon's spin axis. The moon is always treated here as a point mass. This approximation is justified by the timescales at play. As the moon adiabatically follows its Laplace plane equilibrium, its orbital eccentricity remains small at all times (this is verified in all simulations). The only substantial eccentricity increase of the moon is due to crossing the unstable $\mathrm{E}_1$ region. The timescale of exponential eccentricity increase in the $\mathrm{E}_1$ region is given by Eq.~(19) of \cite{saillenfest2021future}, adapted from \cite{tremaine2009satellite}. Once inside the $\mathrm{E}_1$ region, the timescale for the moon's eccentricity to be multiplied by $100$ drops rapidly below hundreds of years for all physical parameters considered here (see also Fig.7 by \citealp{Saillenfest-etal_2022} for a distant Uranus-like planet). This very short timescale should be compared to the eccentricity damping timescale due to tidal dissipation within the moon. Using classical formulas (e.g. Eq.~4.198 by \citealp{Murray-Dermott_1999}) with physical parameters consistent with those of the moons considered in this paper, we invariably obtain timescales that count in millions of years. Due to the many orders of magnitude different between these two timescales, we conclude that tidal dissipation within the moon would not qualitatively change the picture here. The same argument applies to obliquity tides: as the moon adiabatically follows its Laplace plane equilibrium, its spin axis would adiabatically follow its orbit normal. The moon may acquire a substantial non-zero obliquity only in the $\mathrm{E}_1$ region, but there the destabilization timescale is so fast that obliquity tides would not have time to operate.

\section{Slow planet spindown: generating long-lived high-obliquity planets}
\label{sec:longlived}

We first consider the case $\tau_\mathrm{p}\gtrsim\tau_\mathrm{m}$.
\cite{saillenfest2021future} have shown how the tidal migration of a moon can lead to a capture of the host planet in secular spin-orbit resonance. As the moon continues migrating, the planet's obliquity increases and the system converges towards the unstable region $\mathrm{E}_1$. This phenomenon has been shown to be efficient for distant planets such as Jupiter, Saturn, and Uranus, for which the star-driven tidal despinning is negligible (i.e.\ $\tau_\mathrm{p}\gg\tau_\mathrm{m}$). It may also be responsible for the formation of a very oblique ring around the long-period exoplanet HIP\,41378\,f \citep{Saillenfest-etal_2023}. The question then naturally arises about how this mechanism is modified for closer-in planets, for which the star-planet tidal dissipation is non-negligible (i.e. $\tau_\mathrm{p}\gtrsim\tau_\mathrm{m}$). We illustrate this regime with numerical experiments using the code described in Sect~\ref{sec:numcode}.

The interior structure of the planet is unknown, and the presence of resonances with internal oscillations modes can greatly enhance or decrease tidal dissipation (see e.g. \citealp{fuller2016resonance,Farhat-etal_2022}); for this reason, the details of the moon migration law are essentially unconstrained. Therefore, in this section we set the outward orbital migration of the moon as a predefined function of time (e.g. a linear law) and we neglect its corresponding contribution to the despinning of the planet (i.e. we consider that $\tau_\mathrm{p,m}=\infty$). This approximation is relevant for small moons around rapidly rotating planets -- for which the moon orbital angular momentum is negligible compared to the spin angular momentum of the planet. The validity of this approximation is verified a posteriori: including the moon-driven tidal dissipation does not noticeably affect the trajectories shown in this section. The moon is initialized with $10^{-4}$ seed eccentricity and an inclination of $10^{-4}$~rad with respect to its local Laplace plane.

\begin{figure*}
    \centering
    \includegraphics[width=\textwidth]{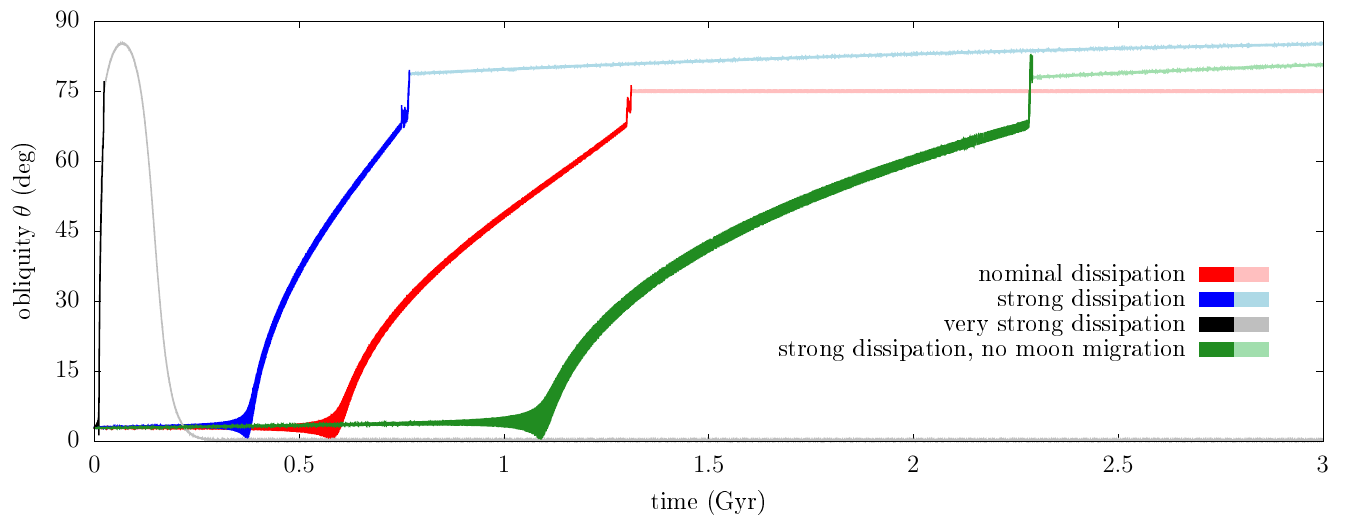}
    \caption{Example of obliquity evolution of the planet HIP\,41378\,f and a hypothetical moon. The mass of the moon is $m/M = 7\times10^{-4}$, and it is initialised at a distance $a_\mathrm{m}=5$~$R_\mathrm{p}$. For the red, blue, and black curves, the lunar migration rate is analogous to that measured for Io and Jupiter \citep{Lainey-etal_2009}. Dark colors show the trajectory when the moon is still present; light colors show the trajectory after the loss of the moon, assumed to be instantly disrupted once its pericenter goes below the Roche limit. For the red curve, the star-planet tidal dissipation is taken into account with tidal time-lag $\Delta t = 3\times 10^{-9}$~years. For the blue curve, the star-planet tidal dissipation is artificially increased using $\Delta t = 2\times 10^{-4}$~years. For the black curve, the star-planet tidal dissipation is artificially increased using $\Delta t = 2\times 10^{-2}$~years. For the green curve, the moon's migration is turned off, and the star-planet tidal dissipation is taken into account using $\Delta t = 2\times 10^{-4}$~years.}
    \label{fig:HIP41378f}
\end{figure*}

The red curve in Fig.~\ref{fig:HIP41378f} shows the obliquity evolution of the exoplanet HIP\,41378\,f as a result of the outward tidal migration of a hypothetical moon. Once the system reaches the unstable region $\mathrm{E}_1$, the moon's eccentricity increases very rapidly, until the moons is disrupted below the planet's Roche limit. The planet's obliquity then becomes extremely stable, away from any resonance (see \citealp{Saillenfest-etal_2023}). In order to power the moon migration, the tidal time lag of the planet is assumed here to be of the order of $\Delta t = 3\times 10^{-9}\;\mathrm{yr}$, which is the value measured for Io and Jupiter\footnote{More precisely, the linear law assumed here for the moon's semi-major axis $a_\mathrm{m}$ produces the same variation range $\Delta a_\mathrm{m}$ during the integration time as a constant time-lag prescription with $\Delta t = 3\times 10^{-9}\;\mathrm{yr}$ (see e.g. \citealp{Efroimsky-Lainey_2007} for the explicit formula).} \citep{Lainey-etal_2009}. HIP\,41378\,f being distant from its host star (period $541$~days; semi-major axis $1.41$~au), the inclusion of the star-planet tidal dissipation with this value of $\Delta t$ does not noticeably affect the dynamics. In other words, $\tau_\mathrm{p}\gg\tau_\mathrm{m}$, so the red trajectory in Fig.~\ref{fig:HIP41378f} is identical to that of \cite{Saillenfest-etal_2023}, Fig.~4.

If we artificially increase the star-planet tidal dissipation, however, the dynamics of HIP\,41378\,f appreciably changes. The blue curve in Fig.~\ref{fig:HIP41378f} shows the evolution when using $\Delta t = 2\times 10^{-4}$~years, which mimics the tidal dissipation felt by a closer-in planet with orbital period $35$~days (i.e. semi-major axis $0.23$~au). Interestingly, resonance capture and obliquity increase still occur, but in a shorter time span. This is mostly due to the gradual spin down of the planet, which decreases its oblateness (see Eq.~\ref{eq:J2}). The planet's spin down consequently decreases the critical distance $r_\mathrm{M}$ that the moon must reach in order to fully tilt the planet through the resonance (see Eq.~\ref{eq:rM_sync}). The tilting mechanism considered by \cite{Saillenfest-etal_2023} is therefore not hampered by star-planet tidal dissipation; it is accelerated. After the disruption of the moon, the planet's obliquity is damped by tidal dissipation, as expected, and the planet converges towards spin-orbit synchronisation and zero obliquity. However, as the planet's obliquity is very large just after the loss of the moon, the effect of tidal dissipation is minimized (see Eqs.~\ref{eq:domegatides} and~\ref{eq:dthetatides}). Hence, for a moderately close-in planet with $\tau_\mathrm{p}\gtrsim\tau_\mathrm{m}$, as in the case shown by the blue curve, the damping process takes tens of billions of years, such that in practice, the planet keeps a large obliquity during the remaining lifetime of the planetary system.

For an even closer-in planet, the resonance capture and obliquity increase due to the moon is only limited to a small transient stage of the evolution (see black curve in Fig.~\ref{fig:HIP41378f}, which mimics a planet with orbital period $10$~days, i.e. semi-major axis $0.1$~au). In that case, we have now $\tau_\mathrm{p}<\tau_\mathrm{m}$. The star-planet tidal dissipation is so strong that the obliquity is rapidly damped after the moon's disruption. The same outcome would have been obtained if the planet had had no moon in the first place and had experienced no obliquity increase. Yet, it is interesting to note that even in the case of a strong dissipation, the resonance capture and obliquity increase still occurs, even though the moon has not much time to migrate at all before the planet has completely spun down. This shows that the star-planet tidal dissipation alone can reproduce the mechanism described by \cite{Saillenfest-etal_2023}, even for a non-migrating moon. This property is illustrated by the green curve in Fig.~\ref{fig:HIP41378f}, for which the moon migration is artificially turned off (i.e.\ $\tau_\mathrm{m}=\infty$). In this example, the resonance capture and obliquity increase is still due to the presence of the moon; however, the increase in $a_\mathrm{m}/r_\mathrm{M}$ is not due to an increase in $a_\mathrm{m}$, but to a decrease in $r_\mathrm{M}$ as a result of the planet's spin-down. This is yet another way of showing that star-planet tidal dissipation assists the moon-induced tilting of the planet described by \cite{Saillenfest-etal_2023}.

\begin{figure}
    \centering
    \includegraphics[width=\columnwidth]{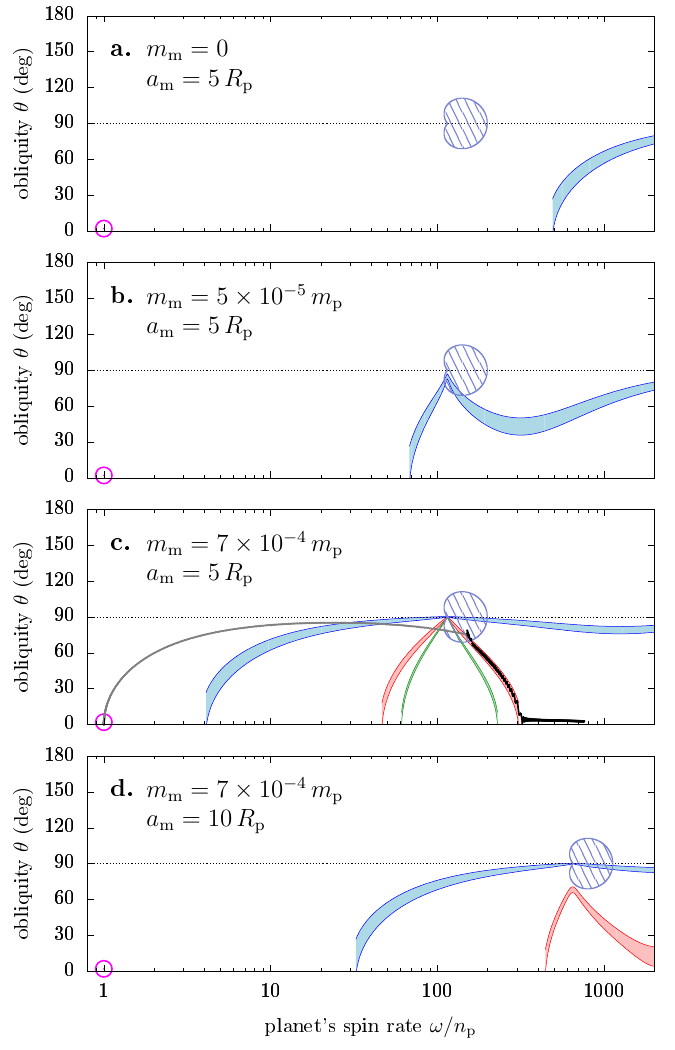}
    \caption{Locations and widths of secular spin-orbit resonances for the exoplanet HIP\,41378\,f for different parameters of a hypothetical moon. The moon is placed on its Laplace equilibrium plane at a fixed distance from the planet (see labels). The three resonances reachable appear in blue, red, and green. They correspond to the resonances labelled $s_3$, $s_6$, and $s_1$ by \cite{Saillenfest-etal_2023}. The hatched blue zone is the $\mathrm{E}_1$ region where the classic Laplace plane of the moon is unstable. The magenta circle highlights the fully relaxed state of the planet ($\omega=n_\mathrm{p}$ and $\theta\approx 0$). Panel~c additionnaly shows the black trajectory from Fig.~\ref{fig:HIP41378f}: the system evolves from right to left; the moon is lost in the $\mathrm{E}_1$ region, then the planet fully relaxes due to tidal dissipation.}
    \label{fig:geometryHIP}
\end{figure}

Previous articles showed how the geometry of resonances depends on the distance $a_\mathrm{m}$ of the moon (see e.g. \citealp{saillenfest2021future,Saillenfest-etal_2023}). If instead, we assume that the moon does not migrate (i.e. $a_\mathrm{m}$ is fixed), but that the planet's spin rate $\omega$ varies as a result of star-planet tidal dissipation, the properties of secular spin-orbit resonances can again be expressed as a function of only two variables: $\omega$ and $\theta$. Figure~\ref{fig:geometryHIP} illustrates the result obtained for HIP\,41378\,f assuming different parameters for the moon. Interesting geometries can be observed. Even though the boost $\alpha'/\alpha$ in the planet's spin-axis precession rate due to the moon varies slowly when $a_{\rm m} \gg r_{\rm M}$ (see Sect.~\ref{ssec:moonsync}), resonances still converge towards a singular point with $\theta=90^\circ$ near $a_{\rm m} \simeq r_{\rm M}$ (the blue hatched region in Panels~c and d). Geometries also depend on the distance $a_\mathrm{m}$ of the moon, taken here as a parameter. Strictly speaking, the geometries observed in Fig.~\ref{fig:geometryHIP} are therefore relevant when the timescale of the star-driven tidal dissipation is much shorter than the timescale of moon migration (i.e. $\tau_\mathrm{p}\ll\tau_\mathrm{m}$): this regime is the focus of the next two sections.

\section{Rapid planet spindown: moon loss through the $\mathrm{E}_1$ instability}\label{sec:e1_instab}

We next turn our attention to to the regime where the planet's spindown is rapid, i.e.\ $\tau_\mathrm{p} \ll \tau_\mathrm{m}$. In this regime, we identify a novel mechanism by which exoplanets can lose their
moons as they spin down.

In previous studies, whether a moon can exist around a given planet is
determined by whether the moon either experiences runaway tidal inspiral or
tidally migrates beyond the stability limit $a_{\rm m} \gtrsim 0.36\, r_{\rm H}$ \citep{barnes2002_moonstab,
sasaki2012_mooninstab}, where $r_{\rm H}$ is the planet's Hill radius and is
given by
\begin{align}
    \frac{r_{\rm H}}{R_{\rm p}} &\equiv \frac{a_{\rm p}}{R_{\rm p}}
            \p{\frac{m_{\rm p}}{3m_\star}}^{1/3},\\
        &\approx 100
            \p{\frac{a_{\rm p}}{0.4\;\mathrm{au}}}
            \p{\frac{R_{\rm p}}{2R_\oplus}}^{-1}
            \p{\frac{m_{\rm p}}{10 M_\oplus}}^{1/3}
            \p{\frac{m_\star}{M_\odot}}^{-1/3}.
            \label{eq:Rhill}
\end{align}
Here, we show that moons need not reach the orbital stability limit to
experience dynamical instability.
The key to our mechanism is the dynamical instability of the $\mathrm{P}_1$ (or $\mathrm{P}_{-1}$) Laplace
equilibrium when the system configuration is within the region $\mathrm{E}_1$. This can drive lunar instability at much smaller orbital separations than in previous works, since
\begin{align}
    \frac{r_{\rm M}}{R_{\rm p}}
        ={}&
            14.2
            \p{\frac{\omega / \omega_{\rm crit}}{0.4}}^{2/5}
            \p{\frac{m_{\rm p}}{10M_\oplus}}^{1/5}
            \p{\frac{R_{\rm p}}{2R_\oplus}}^{-3/5}\nonumber\\
        &\times
            \p{\frac{a_{\rm p}}{0.4\;\mathrm{au}}}^{3/5}
            \p{\frac{m_\star}{M_{\odot}}}^{-1/5}.\label{eq:rM_Rp}
\end{align}
In order for the moon to experience this dynamical instability, the system must
evolve through $\mathrm{E}_1$. There are two distinct tidal dissipation mechanisms that can contribute to such
an evolution, which correspond to the limits $\tau_\mathrm{p}\ll\tau_\mathrm{m}$ and $\tau_\mathrm{p}\gg\tau_\mathrm{m}$.

First, for $\tau_\mathrm{p}\ll\tau_\mathrm{m}$, the tidal evolution of the young planet's spin (predominantly driven by
the host star) naturally acts to drive systems towards $\mathrm{E}_1$.
As the planet spins down, its $r_{\rm M} \propto \omega^{2/5}$ decreases
(see Eq.~\ref{eq:rM_sync}).
Thus, since initially $a_{\rm m} < r_{\rm M}$, the equality $a_{\rm m} = r_{\rm M}$ is satisfied at some point during the planet's spindown.
At the same time, tidal dissipation also drives the young planet's obliquity
towards $90^\circ$ due to tidal dissipation when the planet's spin is
sufficiently rapid: to be precise, $\cos\theta$ is driven towards $2n_{\rm p} /
\omega \ll 1$, see Eq.~\eqref{eq:dthetatides}\footnote{
We note that while we consider only the constant time lag model here (see Sect.~\ref{ssec:dissip}), more general conditions for tidal obliquity excitation can be developed for general equilibrium tidal models \citep[e.g.][]{Boue-etal_2016, valente2022tidal}, and obliquity excitation can even arise in dynamical tidal models such as resonance locks \citep[e.g. see appendix of][]{zanazzi2024damping}.}.
Combining these two observations, it is clear that evolution through
$\mathrm{E}_1$ can
be very common.
We note that a passage through $\mathrm{E}_1$ leads to dynamical instability of the moon's orbit on
short timescales ($\sim n_{\rm m} / n_{\rm p}^2$, which is much shorter than the tidal timescale
$\tau_{\rm p}$; see Eq.~9 of
\citealp{saillenfest2021future}).

\begin{figure}
    \centering
    \includegraphics[width=\columnwidth]{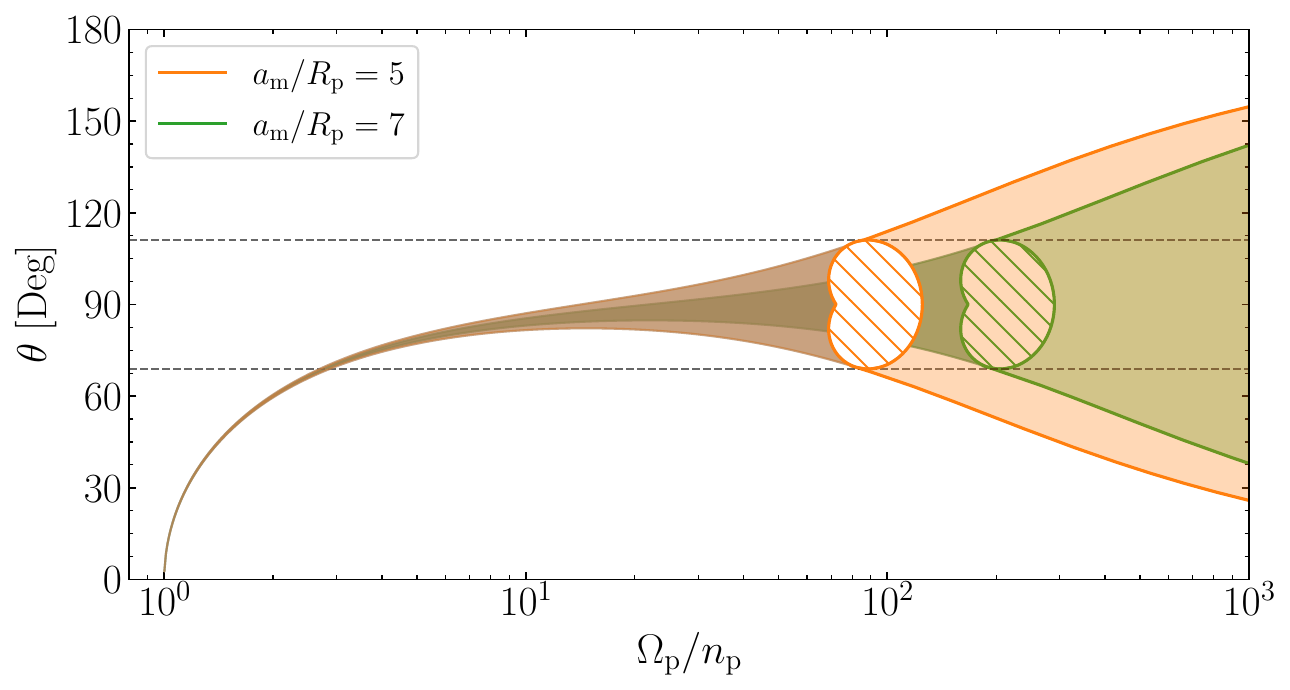}
    \caption{
    Plot of the $\mathrm{E}_1$ region for two different values of $a_{\rm m} / R_{\rm
    p}$.
    For a planet born rapidly rotating, no moon will be found in the
    shaded regions of parameter space, which will pass through the
    $\mathrm{E}_1$ region
    during the planet's tidal despinning and alignment.
    The $\mathrm{E}_1$ region can also be encountered due to lunar migration if the lunar
    migration timescale is faster than the planet's despinning timescale.
    In this case, the moon is lost if the planet's obliquity lies within about $69^\circ$ and $111^\circ$ (i.e.\ between the black dashed lines) independent of $a_{\rm m} / R_{\rm p}$.
    }\label{fig:instab}
\end{figure}

In Fig.~\ref{fig:instab}, the lightly shaded regions denote the regions of
parameter space such that the planet's tidal evolution results in passage through $\mathrm{E}_1$; we have assumed that (i) the moon's distance does not vary, and (ii) the moon adiabatically follows its Laplace state as the planet spins down.
We consider two lunar distances, $a_\mathrm{m}=5$~$R_\mathrm{p}$ and $7$~$R_\mathrm{p}$, consistent with the locations of major regular moons in the Solar System.
We note that for sufficiently small ($\theta \approx 0^\circ$) or large ($\theta
\approx 180^\circ$) initial obliquities, the instability is not reached, but it is
reached for a broad range of intermediate obliquities.
This range is wider for planets that are born with larger $\omega / n_{\rm p}$, that is, for more distant planets that must despin more. However, the tidal timescale $\tau_\mathrm{p}$ of more distant planets is larger, and possibly comparable with the moon migration timescale $\tau_\mathrm{m}$, which leads us to the second case.

If instead $\tau_\mathrm{p}\gg\tau_\mathrm{m}$, the system can \emph{still} pass through the $\mathrm{E}_1$ region because of the tidal evolution of the lunar orbit alone.
In this case, the lunar orbit widens
(due to the tide it raises on the rapidly rotating planet) such that $a_{\rm m}$
approaches $r_{\rm M}$ from below.
Such an evolution results in $\mathrm{E}_1$ passage for an intermediate range of
obliquities (denoted by the dashed curves in
Fig.~\ref{fig:instab}). As seen in Fig.~\ref{fig:instab}, the fraction of initial conditions that result in $\mathrm{E}_1$ passage and lunar
dynamical instability is smaller when $\tau_\mathrm{p}\gg\tau_\mathrm{m}$ than when $\tau_\mathrm{p}\ll\tau_\mathrm{m}$. Intermediate regimes for which $\tau_\mathrm{p}\approx\tau_\mathrm{m}$ would lie between these two extreme cases.

We note that in Fig.~\ref{fig:instab}, we assume that the planet does not encounter any secular spin-orbit resonance. If secular spin-orbit resonances are present due to other planets in the system, the planet's spin down and/or moon migration may indirectly induce obliquity changes for the planet, and possibly drive the system towards $\mathrm{E}_1$ (see Sect.~\ref{sec:longlived}). Depending on the precise location of resonances, the existence of resonances may therefore increase the fraction of initial conditions that result in lunar instability.

\begin{figure}
    \centering
    \includegraphics[width=\linewidth]{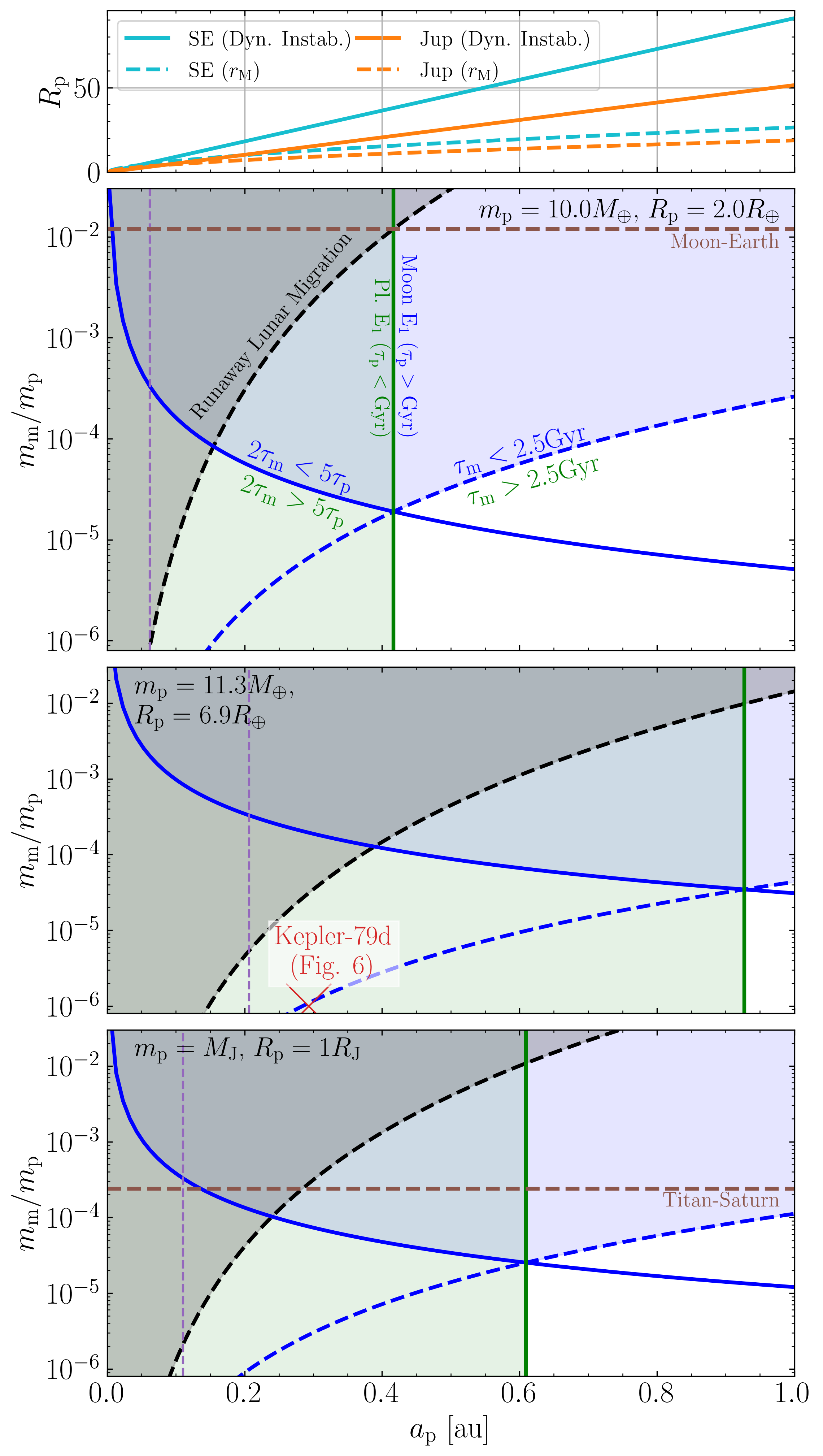}
    \caption{
    The different mechanisms of moon loss as a function of the system's parameters.
    The top panel shows the instability limit ($a_{\rm m} = 0.36\,r_{\rm H}$) for a
    Super Earth-like and Jupiter-like planet in solid lines, and the midpoint
    radius $r_{\rm M}$ (evaluated with the planet spinning at $2/3$ its maximal
    rate) in dashed lines.
    The second panel shows the regions of parameter space for the lunar
    mass ratio ($m_{\rm m} / m_{\rm p}$) and planet semi-major axis ($a_{\rm
    p}$) for a typical super Earth (SE).
    The grey region above the black dashed line is where moons are lost anyway through runaway migration \citep{barnes2002_moonstab}.
    The vertical green line denotes the boundary between where planets do and do not despin within $\sim 1$~Gyr (Eq.~\ref{eq:taup}).
    The blue solid curve denotes the boundary where lunar tidal migration (Eq.~\ref{eq:taum}) and tidal planet despinning result in comparable changes to $a_{\rm m} / r_{\rm M}$ (Eq.~\ref{eq:E1_ratio}): above this curve, $\mathrm{E}_1$ is crossed due to lunar migration; below this curve, $\mathrm{E}_1$ is crossed due to planetary despinning.
    The blue dashed line denotes the boundary where lunar tidal migration results in significant changes to $a_{\rm m} / r_{\rm M}$ in $\sim$~Gyr timescales.
    Accordingly, the region where moons are lost due to lunar migration (with or without capture in secular spin-orbit resonance) is shaded blue, and the region where moons are lost due to planetary despinning is shaded green.
    The vertical purple line denotes the point at which $r_{\rm M} = 5R_{\rm p}$ (Eq.~\ref{eq:rM_Rp}).
    The middle panel is the same for the parameters of Kepler-79d, and the red cross denotes the fiducial parameters adopted for a hypothetical
    moon (see Fig.~\ref{fig:Kepler79d}).
    The bottom panel is the same for a Jupiter-mass planet.
    }\label{fig:barnes_comp}
\end{figure}

In Fig.~\ref{fig:barnes_comp}, we show the regions of parameter space where each
regime dominates.
\cite{barnes2002_moonstab} show that moons cannot be found
around planets if their maximum lunar migration time, found when the moon is
located at $0.36\,r_{\rm H}$, is $\lesssim \;\mathrm{Gyr}$.
We can evaluate this region of parameter space using Eq.~\eqref{eq:taum}; it
is shown as the grey shaded region in the bottom three panels of
Fig.~\ref{fig:barnes_comp}.
In our mechanism, we require that either the planet's despinning time $\tau_\mathrm{p}$
(Eq.~\ref{eq:taup}) or the lunar migration time $\tau_\mathrm{m}$ evaluated at the edge of
$\mathrm{E}_1$
(namely, $r_{\rm M} / 3^{1/5}$; see \citealp{saillenfest2021future}) be $\lesssim
\;\mathrm{Gyr}$.
These two conditions correspond to the green and blue shaded regions of
parameter space respectively.
It can be seen for common lunar mass ratios that our proposed mechanism greatly
expands the region of parameter space where exomoons are lost.

In Fig.~\ref{fig:barnes_comp}, we have taken care to distinguish between the regions where the $\mathrm{E}_1$ instability is crossed due to planetary spindown (green shaded regions) or to lunar migration (blue shaded regions).
The exact mechanism responsible changes the fraction of initial
conditions susceptible to dynamical instability (see Fig.~\ref{fig:instab}).
To quantify the boundary between these two regimes, we compare the contributions
of lunar migration ($\dot{a}_{\rm m}$) and planetary spindown ($\dot{\omega}$)
to the time evolution of the ratio $a_{\rm m} / r_{\rm M}$.
Quantitatively, the two contributions are equal when $\kappa=1$, where we define
\begin{align}
    \kappa ={}& \frac{5(\mathrm{d}\ln a_{\rm m} / \mathrm{d}t)_{a_{\rm m} = r_{\rm M} / 3^{1/5}}}
        {2(\mathrm{d}\ln \omega / \mathrm{d}t)_{\omega = \omega_0}}
        \approx
            \frac{45 m_{\rm m}}{4k_2m_{\rm p}}
            \frac{\omega_{\rm crit}^2}{n_{\rm p}\omega_0}\nonumber\\
        \approx{}& 20
            \p{\frac{m_{\rm m} / m_{\rm p}}{10^{-4}}}
            \p{\frac{\omega_0}{2\pi / (7.5\;\mathrm{hr})}}^{-1}
            \p{\frac{\rho}{\rho_{\oplus}}}\nonumber\\
            &\times \p{\frac{m_\star}{M_{\odot}}}^{-1/2}
            \p{\frac{a_{\rm p}}{0.29\;\mathrm{au}}}^{-3/2}.\label{eq:E1_ratio}
\end{align}
When $\kappa < 1$, the planet's spin down is the main driver of variations in $a_\mathrm{m}/r_\mathrm{M}$ resulting in crossing the $\mathrm{E}_1$ region. In that case, a substantial tidal evolution of the planetary obliquity is expected (Fig.~\ref{fig:instab}). In contrast, when $\kappa > 1$, the moon migration is the main driver of variations in $a_\mathrm{m}/r_\mathrm{M}$. In that case, the moon reaches the $\mathrm{E}_1$ instability before the planet
obliquity has any chance to change, so moons are lost only for planets with initial obliquities
compatible with the $\mathrm{E}_1$ range (approximately $\in [69^\circ, 111^\circ]$).
We see that efficient lunar ejection via the $\mathrm{E}_1$ instability favors lower-mass moons and less-dense planets.
Fortuitously, young planets are expected to be less dense \citep[e.g.][]{thao2024_featherweight}.

\begin{figure*}
    \centering
    \includegraphics[width=0.8\textwidth]{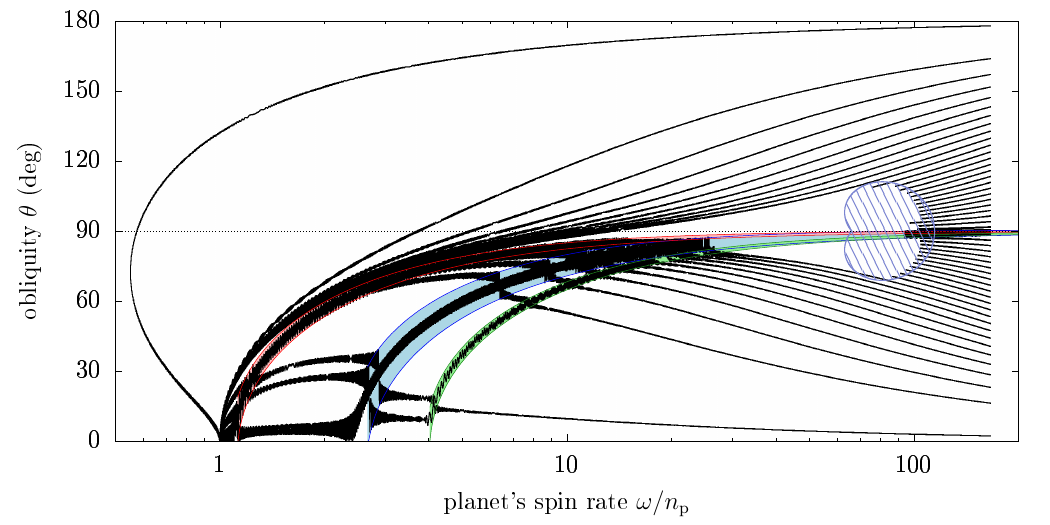}
    \caption{
    Self-consistent obliquity and spin rate evolution for several realizations of exoplanet
    Kepler-79\,d.
    Trajectories go from right to left.
    The planet is initalized with a spin rate of $7.5$~hours (i.e.\
    $\omega/n_\mathrm{p}\approx 170$) and different obliquities equispaced in
    $\cos\theta$. A moon with mass $m_\mathrm{m}/m_\mathrm{p}=10^{-6}$ is initialized at a distance $a_\mathrm{m}=4$~$R_\mathrm{p}$.
    Integrations are stopped when the moon's pericenter goes below the Roche limit.
    The three largest resonances reacheable by Kepler-79\,d appear in red, blue,
    and green.
    The hatched blue zone is the $\mathrm{E}_1$ region.
    For definitiveness, the resonances and $\mathrm{E}_1$ region in the background are drawn assuming that the moon
    is on a circular orbit at $a_\mathrm{m}=4$~$R_\mathrm{p}$; for moons that do not cross the $\mathrm{E}_1$ region, this remains approximatively true during the whole simulation.
    }\label{fig:Kepler79d}
\end{figure*}

We now turn to numerical simulations of this mechanism using the same secular code as previously (see Sect.~\ref{sec:numcode}). Figure~\ref{fig:Kepler79d} illustrates the moon disruption process due to planetary spin down for an existing exoplanet. We consider the exoplanet Kepler-79\,d, which has a period of $52$~days and is member of a four-planet system (see \citealp{Jontof-Hutter-etal_2014,Yoffe-etal_2021}). Kepler-79\,d belongs to the class of `super-puff' exoplanets, with its measured density being $0.2$~$\mathrm{g}\,\mathrm{cm}^{-3}$. The large radius of Kepler-79\,d is still unexplained; however, its present-day parameters are coincidentally consistent with those of a young, inflated sub-Neptune, so for concreteness we adopt its present-day mass and radius: $M = 11.3$~$M_{\oplus}$ and $R = 6.912$~$R_{\oplus}$ \citep{Yoffe-etal_2021}. The three other planets in the system generate secular spin-orbit resonances that Kepler-79\,d can encounter during its tidal evolution. The tidal time lag $\Delta t$ of Kepler-79\,d is assumed to be $10^{-6}$~years for all simulations. The moon migration and tidal despinning of the planet, both due to stellar and lunar tides, are included self consistently. The moon migrates as a response of angular momentum transfer. It is initialized on its Laplace surface (in $\mathrm{P}_1$ for $\theta<90^\circ$, $\mathrm{P}_{-1}$ for $\theta>90^\circ$) with seed eccentricity $e_\mathrm{m}=10^{-4}$.

In Fig.~\ref{fig:Kepler79d}, it can be seen that for intermediate planetary obliquities the moon becomes dynamically unstable, and is eventually tidally disrupted as its eccentricity grows, when it encounters the $\mathrm{E}_1$ region. This behaviour was predicted analytically in Figs.~\ref{fig:instab} and \ref{fig:barnes_comp}. We note that the moon is lost slightly after encountering the limit of the $\mathrm{E}_1$ region; this is due to the fact that the moon first tranfers to an eccentric stable equilibrium, which then becomes unstable, too (see \citealp{tremaine2009satellite,saillenfest2021future}). For trajectories that do not encounter the $\mathrm{E}_1$ region, the moon survives and the system continues to evolve due to tidal dissipation. Most trajectories then cross one or several secular spin-orbit resonances; some trajectories are captured in resonance and follow the resonance center (Cassini State 2) as the planet despins. When trajectories cross a resonance separatrix, the capture process can be modelled as a probabilistic event (see e.g. \citealp{su2021dynamics}).

In Fig.~\ref{fig:Kepler79d}, we intentionally adopt a very small moon mass in order for the system to be firmly in the green region of Fig.~\ref{fig:barnes_comp} (i.e. $\kappa\ll 1$ in Eq.~\ref{eq:E1_ratio}). With such a small moon mass, $\tau_\mathrm{p}$ is significantly shorter than $\tau_\mathrm{m}$, which means that the moon only slightly migrates during the entire simulation; as a result, the crossing of the $\mathrm{E}_1$ region is largely due to the planetary spin down. For larger moon masses (not shown), the $\mathrm{E}_1$ region would be reached earlier because the moon migration would be more substantial.

\section{Moons and spin-axis dynamics of despun planets}\label{sec:despun}

In this section, we examine the effect of lunar migration on the spin evolution of tidally despun planets, that is, when the planet is close to spin-orbit synchronization.
This corresponds to the late-time regime of the same $\tau_\mathrm{p} \ll \tau_{\mathrm{m}}$ limit considered in Section~\ref{sec:e1_instab}.
While lunar migration can have dramatic effects on the obliquities of distant fast-spinning planets \citep[e.g.][]{Saillenfest-etal_2021, Saillenfest-etal_2023}, its impact on close-in despun planets remains to be investigated.

\subsection{Obliquity equilibria broken by moon inspiral}\label{ssec:moon_tcebreak_numerical}

Around slowly spinning planets, moons all migrate inwards due to tidal dissipation -- irrespective of whether the moon is prograde or retrograde. When the moon is sufficiently close to the planet, the tide it raises on the planet dominates
the tide raised by the star.
By comparison of Eqs.~\eqref{eq:taup} and~\eqref{eq:tau_pm}, we find that this
occurs when the lunar semi-major axis is smaller than the critical value $a_{\rm
m, crit}$ satisfying:
\begin{align}
            \p{\frac{m_\star}{m_{\rm p}}}
            \p{\frac{R_{\rm p}}{a_{\rm p}}}^3
            n_{\rm p}
        &=
            \p{\frac{m_{\rm m}}{m_{\rm p}}}
            \p{\frac{R_{\rm p}}{a_{\rm m, crit}}}^3
            n_{\rm m},\\
\frac{a_{\rm m, crit}}{a_{\rm p}}
    &=
        \p{\frac{m_{\rm m}}{\scinot{3}{-4}m_{\rm p}}}^{2/9}
        \p{\frac{m_{\rm p}}{10 M_\oplus}}^{1/3}
        \p{\frac{m_{\star}}{M_\odot}}^{-1/3}
            .\label{eq:am_crit}
\end{align}
When the moon is at a distance $a_\mathrm{m,crit}$ from the planet, its orbital angular momentum $L_\mathrm{m}$ can quite easily be comparable to the spin angular momentum $S_{\rm p}$ of the despun planet:
\begin{align}
    \frac{S_{\rm p}}{L_{\rm m, crit}}
        ={}& \frac{k_{\rm p}m_{\rm p}R_{\rm p}^2\omega}{
            m_{\rm m}\sqrt{Gm_{\rm p}a_{\rm m, crit}}}\nonumber\\
        &= 0.12
            \p{\frac{m_\star}{M_{\odot}}}^{2/3}
            \p{\frac{m_{\rm p}}{10M_{\oplus}}}^{-2/3}
            \p{\frac{m_{\rm m} / m_{\rm p}}{\scinot{3}{-4}}}^{-10/9}\nonumber\\
        &\times \p{\frac{R_{\rm p}}{2R_{\oplus}}}^{2}
            \p{\frac{a_{\rm p}}{0.4\;\mathrm{AU}}}^{-2}.
    \label{eq:angmom_ratio_crit}
\end{align}
As such, as the moon inspirals, the spin of the planet tends to synchronize with the orbit of its moon.
The runaway lunar inspiral stops either when the moon is tidally disrupted below the Roche limit, or when the planet becomes tidally synchronized to the moon.
For typical lunar mass ratios ($m_{\rm m} \lesssim 10^{-3}m_{\rm p}$), the
latter is impossible, as the moon's orbit does not contain enough angular momentum to spin up the planet before being disrupted (the classic Darwin instability; \citealp{darwin1879_instab, 2023makarov_survival}).
Nevertheless, the migration of moons can still appreciably change the final spin states of
planets.

As discussed in Section~\ref{sec:theory}, a moon-less planet in a multiplanetary
system tidally evolves towards one of two tidal Cassini Equilibria
\citep{levrard2007, su2021dynamics}, one at low obliquity (tCE$_1$) and one at high obliquity
(tCE$_2$).
After the planet has reached one of these tCE due to the stellar tide, we now consider the effect of the runaway lunar inspiral.
If a prograde moon undergoes runaway inspiral around a planet in tCE$_1$
($\theta\approx 0^\circ$ and $\omega/n_\mathrm{p}\approx 1$), then the moon
simply spins the planet up for a short amount of time without changing its
obliquity.
Then, after the moon has been disrupted below the Roche limit, the planet
converges back to $\omega/n_\mathrm{p}=1$.

On the other hand, if runaway lunar inspiral occurs around a planet in tCE$_2$
(high obliquity, subsynchronous spin), both the spin rate and obliquity of the
planet can be affected.
We expect that the moon's tidal torque, which grows as it inspirals, rapidly
overwhelms the spin-orbit resonance (see Sect.~\ref{ssec:tce_break}), with the result of ejecting the planet out of tCE$_2$.
In such cases, the planet's spin then aligns with the lunar orbit.
Eventually, after the moon is disrupted below the Roche limit, the planet tidally evolves to
another equilibrium state.

In order to investigate this scenario, we ran additional simulations with the model of Sect.~\ref{sec:numcode}, still taking
into account the tidal dissipation due to both the star and the moon
self-consistently following the constant time-lag theory.
We adopt the following parameters: the star has mass $m_\star = 1$~$M_\odot$,
the planet has mass $m_\mathrm{p}=10$~$M_\oplus$, semi-major axis
$a_\mathrm{p}=0.4$~au, and zero eccentricity.
The moon has mass $m_\mathrm{m}/m_\mathrm{p} = 3\times 10^{-4}$; it is
initialised on its Laplace plane (see below) with $10^{-4}$ seed eccentricity.
The planet's Love number is set to $k_2=0.29$ and constant time-lag $\Delta t =
7\times 10^{-5}$~years.
The system is perturbed by a $1$~$M_\mathrm{J}$ planet at $10$~au inclined by
$10^\circ$, which generates a large secular spin-orbit resonance.
In order to produce a (probabilistic) capture into resonance, the planet is
initialized with a retrograde obliquity $\theta \approx 150^\circ$ and spin
period $2\pi/\omega\approx 5$~hours.
We note that, for initially retrograde-spinning planets, the lunar orbit in the $\mathrm{P}_1$
Laplace state is retrograde with respect to the spin of the planet
($i_{\rm m1} > 90^\circ$, point S1 in Fig.~\ref{fig:laplace_p1}).
Thus, for completeness, we consider both initially prograde ($\mathrm{P}_{-1}$) and retrograde-orbiting ($\mathrm{P}_1$)
moons as initial conditions.

\begin{figure*}
    \centering
    \includegraphics[width=\textwidth]{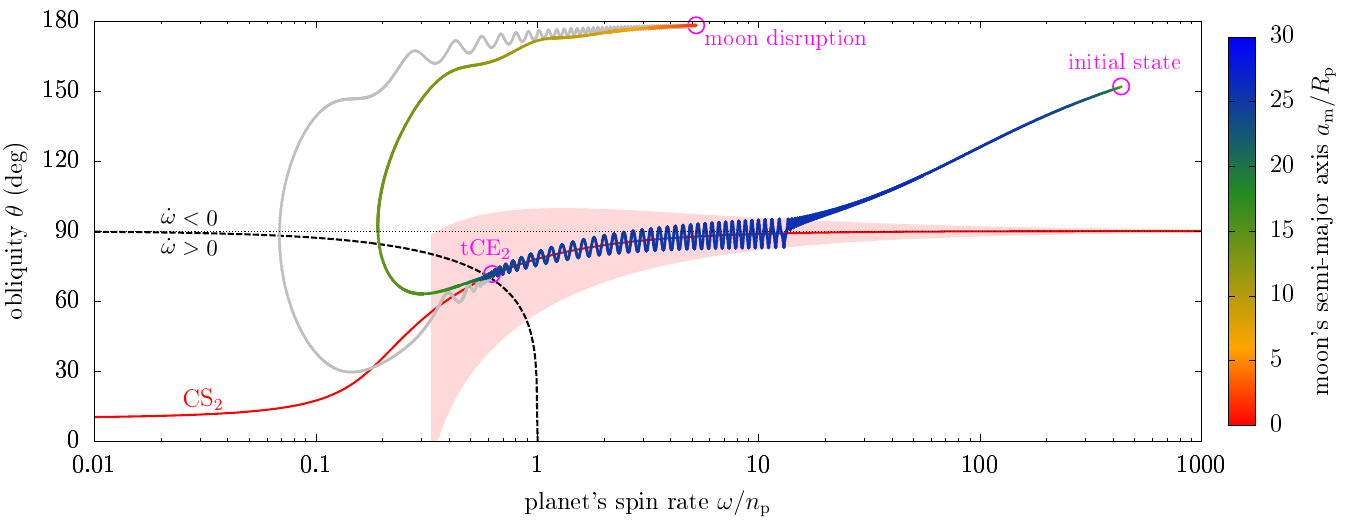}
    \caption{
    Example of ejection from a tidal Cassini Equilibrium due to moon inspiral.
    The color of the curve shows the moon's distance to the planet; the grey
    curve shows the evolution after the moon has been lost.
    The center of the resonance (Cassini state~2; red curve) and its width (pink
    region) are shown at capture time, when the moon's semi-major axis is
    $a_\mathrm{m}\approx 25$~$R_\mathrm{p}$.
    Along the dashed black curve, the star-driven tidal dissipation results in
    $\dot{\omega}=0$ (see Eq.~\eqref{eq:domegatides}).
    }\label{fig:tCE2break_retromoon}
\end{figure*}

First, we consider the case in which the orbit of the moon is initially prograde
with respect to the planet's spin; this would be expected if the moon forms from
the same disk of material as does the planet, and corresponds to the Laplace equilibrium $\mathrm{P}_{-1}$.
The resulting evolution is shown in Fig.~\ref{fig:tCE2break_retromoon}.
The moon first migrates outwards, because it is initially located beyond the
synchronous radius.
However, as the spin rate of the planet is still high, the amount of angular
momentum extracted by the moon is negligible.
The tidal evolution of the planet is driven by the stellar tides, which damp the
planet's obliquity and spin rate.
In this example, the planet is captured in the secular spin-orbit resonance, and
it converges towards the tidal Cassini Equilibrium $\mathrm{tCE}_2$, in a similar way to moon-less planets \citep{levrard2007, su2021dynamics}.
During this process, the moon's migration is stalled at $a_\mathrm{m}\approx
25$~$R_\mathrm{p}$.
As the planet's obliquity and oblateness vary, the moon's orbital plane
adiabatically follows the orientation of the Laplace plane $\mathrm{P}_{-1}$ (see
Eq.~\eqref{eq:def_im1prime}).
For about $100$~Myrs, the system remains at $\mathrm{tCE}_2$ with no much
variation, except that the moon slowly inspirals towards the planet.
When the moon reaches a distance of $a_\mathrm{m}\approx 15$~$R_\mathrm{p}$,
however, it starts its runaway inspiral and its angular momentum exchange with
the planet becomes more efficient.
This ejects the planet out of $\mathrm{tCE}_2$.
Because at $\mathrm{tCE}_2$ the moon is now \emph{retrograde} to the planet's
spin axis (see left panel of Fig.~\ref{fig:laplace_p1}),
the moon's runaway inspiral initially spins the planet further down, and it
drives the planet's spin axis towards $\theta\approx 180^\circ$, that is,
towards alignment with the (retrograde) moon's orbital angular momentum.
Yet, once the planet's obliquity goes back to values $\theta > 90^\circ$, the
moon's orbit is prograde again to the planet's spin axis, and the moon's
inspiral spins the planet back up.
At the end of the moon's runaway inspiral, the planet's spin angular momentum
and the moon's orbital angular momentum are approximatively aligned, and the
planet's obliquity is $\theta\approx 180^\circ$.
At that point, the moon reaches the planet's Roche limit (considered here to be
$2$~$R_\mathrm{p}$): we assume that the moon is instantly disrupted into pieces
that collide with each other and efficiently damp the remaining angular momentum
of the moon.
The spin dynamics of the---now moon-less---planet is then driven by stellar
tides only.
In the example of Fig.~\ref{fig:tCE2break_retromoon}, the planet is trapped again in Cassini State 2, and it
eventually converges back to $\mathrm{tCE}_2$.

\begin{figure*}
    \centering
    \includegraphics[width=\textwidth]{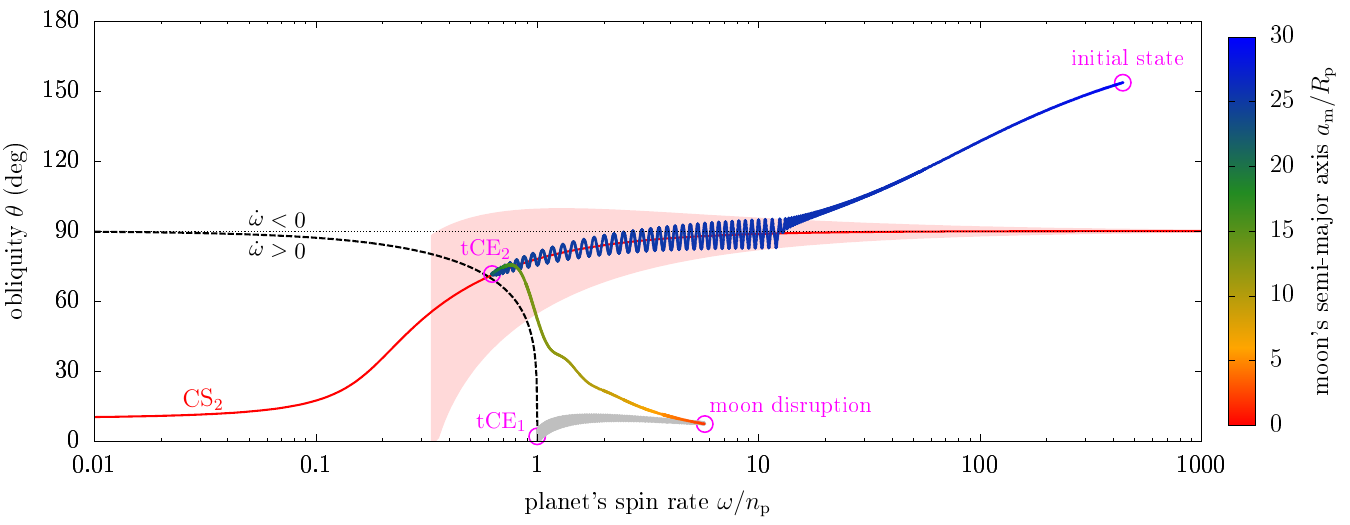}
    \caption{
    Same as Fig.~\ref{fig:tCE2break_retromoon}, but with an initially retrograde
    moon.
    }\label{fig:tCE2break}
\end{figure*}

The peculiar evolution of the system observed here reveals the decisive role
of the moon's angular momentum during its runaway inspiral, despite the modest
mass of the moon $m_\mathrm{m}/m_\mathrm{p} = 3\times 10^{-4}$.
As a variant scenario, we wonder what would be the effect of a prograde
inspiralling moon, which would spin the planet up at $\mathrm{tCE}_2$ (instead
of spinning it down), and align its spin axis towards $\theta\approx 0^\circ$
(instead of $180^\circ$).
The result is shown in Fig.~\ref{fig:tCE2break}.
In order for the moon's orbit to be prograde to the planet's spin axis at
$\mathrm{tCE}_2$, we initialise it at the Laplace equilibrium $\mathrm{P}_1$.
Such a `regular retrograde' moon can be formed as a result of a giant impact
that would have tilted the planet to such a large initial obliquity (see e.g.\ \citealp{Morbidelli-etal_2012}).
Contrary to the previous case, the moon initially migrates inwards---because it
is retrograde; then, it still migrates inwards when the planet has reached
$\mathrm{tCE}_2$---because it is prograde and located below the synchronization
radius.
Again, the system remains at $\mathrm{tCE}_2$ for about $100$~Myrs, while the
moon slowly inspirals.
However, when the moon starts its runaway inspiral, it gives angular momentum to
the planet and strongly torques its spin axis towards $\theta\approx 0^\circ$,
which breaks the resonance.
When the moon reaches the Roche limit, the planet's spin angular momentum and
the moon's orbital angular momentum are approximatively aligned, with the
planet's obliquity now $\theta\approx 0^\circ$.
The stellar tide then directly drives the planet to the tCE$_1$ state, with $\theta\approx 0^\circ$ and $\omega/n_\mathrm{p}\approx
1$. Hence, in this example, the moon's inspiral and disruption has changed the equilibrium spin state of the planet from a high obliquity to a low obliquity.

\begin{figure*}
    \centering
    \includegraphics[width=\textwidth]{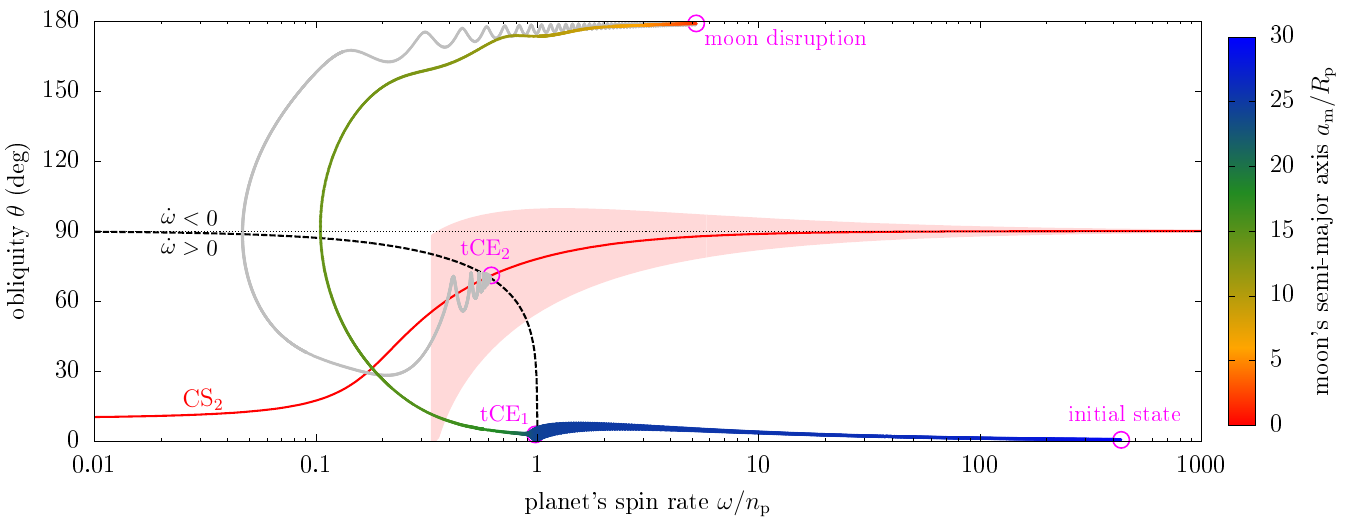}
    \caption{
    Same as Fig.~\ref{fig:tCE2break}, but for an initially low-obliquity planet.
    }\label{fig:tCE1break}
\end{figure*}

The reversed situation can be obtained if the planet initially has a low obliquity---in that case, the existence of a retrograde moon can be explained by a captured object like Triton around Neptune. Figure~\ref{fig:tCE1break} shows that the planet first converges to $\mathrm{tCE}_1$, but it is then ejected by the moon's runaway inspiral, and it eventually stabilizes at $\mathrm{tCE}_2$.

\subsection{Long-lived high obliquity due to moon-driven spin up}\label{ssec:massive_moon_longlived}

Above, we pointed out that moons with typical mass ratios $m_\mathrm{m}/m_\mathrm{p}\lesssim 10^{-3}$ experience runaway
inspiral and are disrupted before they can tidally synchronize their host planets.
However, when the lunar mass is sufficiently large, the moon can considerably spin up and
realign the planet \citep{2023makarov_survival, makarov2025_moonsync}, introducing new possibilities
for spin-orbit resonances.
We consider this possibility here.

As the moon synchronizes the planet, conservation of angular momentum of
the moon-planet system demands that the new spin rate of the planet $\omega_{\rm
new}$ (also equal to the new mean motion of the moon) satisfies
\begin{align}
    m_{\rm m}\sqrt{Gm_{\rm p}a_{\rm m, crit}}
        &\approx \s{
            k_{\rm p}m_{\rm p}R_{\rm p}^2 + m_{\rm m} \p{Gm_{\rm p} /
            \omega_{\rm new}^2}^{2/3}}\omega_{\rm new},\label{eq:om_moonpl}\\
    \sqrt{\frac{a_{\rm m, crit}}{R_{\rm p}}}
        &\approx
            \s{
                k_{\rm p}\frac{m_{\rm p}}{m_{\rm m}}
                + \frac{1}{\tilde{\omega}_{\rm new}^{4/3}}
                }\tilde{\omega}_{\rm new}. \label{eq:f_new}
\end{align}
Here, $\tilde{\omega}_{\rm new} \equiv \omega_{\rm new} / \omega_{\rm crit}$, $a_{\rm m,crit} \simeq 10R_{\rm p}$ (Eq.~\ref{eq:am_crit}), and we have neglected the
contribution of the planet's spin angular momentum to the initial angular
momentum budget (see Eq.~\ref{eq:angmom_ratio_crit}).
A solution for $\tilde{\omega}_{\rm new}$ only exists if the minimum of the right-hand side
is less than the left-hand side, which requires
\citep[cf.][]{2023makarov_survival}
\begin{align}
    \frac{m_{\rm m}}{m_{\rm p}}
        \gtrsim{}&
            \scinot{2}{-3}
                \p{\frac{R_{\rm p}}{2R_{\oplus}}}^{18/13}
                \p{\frac{a_{\rm p}}{0.4\;\mathrm{AU}}}^{-18/13}\nonumber\\
            &\times \p{\frac{m_\star}{M_{\odot}}}^{6/13}
                \p{\frac{m_{\rm p}}{10M_{\oplus}}}^{-6/13}.
                \label{eq:mmoon_min}
\end{align}
Note that this is a significantly more massive moon than used previously in our fiducial
parameters.
For an insufficiently massive moon, the right-hand side of Eq.~\eqref{eq:f_new} always exceeds the left-hand side, and the moon is tidally disrupted by
the planet before it synchronizes the planet's spin.
In contrast, for a sufficiently massive moon, the new planetary spin rate
is given by the smaller root of Eq.~\eqref{eq:f_new}, which yields
\begin{align}
    \tilde{\omega}_{\rm new}
        &\simeq \p{\frac{R_{\rm p}}{a_{\rm m, crit}}}^{3/2},\\
    \frac{\omega_{\rm new}}{n_{\rm p}}
        &\simeq \p{\frac{m_{\rm p}}{m_{\rm m}}}^{1/3}.\label{eq:massivemoon_spunup}
\end{align}
This estimate neglects the solar tide during the moon-driven spinup; when
including the spindown driven by the solar tide, $\omega_{\rm new}$ will be
somewhat lower than this.
Note that this result seems somewhat counter-intuitive at first glance: a more
massive moon spins the planet up less!
There is a simple explanation: for a more massive moon, spinning up the planet
requires a smaller change to the lunar orbit, resulting in a smaller lunar mean
motion (and planetary spin rate).

Once the massive moon has synchronized the planet, the planet-moon system is in
an equilibrium of the lunar tidal evolution.
As such, further evolution of the planet's spin proceeds due to the stellar tide,
which acts to weakly drive the planet back to its orbital frequency.
Since the lunar tidal torque is stronger than the stellar one, the planet's spin
angular momentum lost to the stellar tidal torque is rapidly replenished by the
moon's orbital angular momentum.
Due to the large mass and angular momentum of the moon, this significantly
extends the lifetime of the planet's supersynchronous spin state.
However, once even the moon's angular momentum is depleted by the stellar tide,
the moon is tidally disrupted, and the planet rapidly evolves to a tCE following
the moon-less spin-orbit evolution (see Sects.~\ref{ssec:res}
and~\ref{ssec:dissip}).

\begin{figure*}
    \centering
    \includegraphics[width=\textwidth]{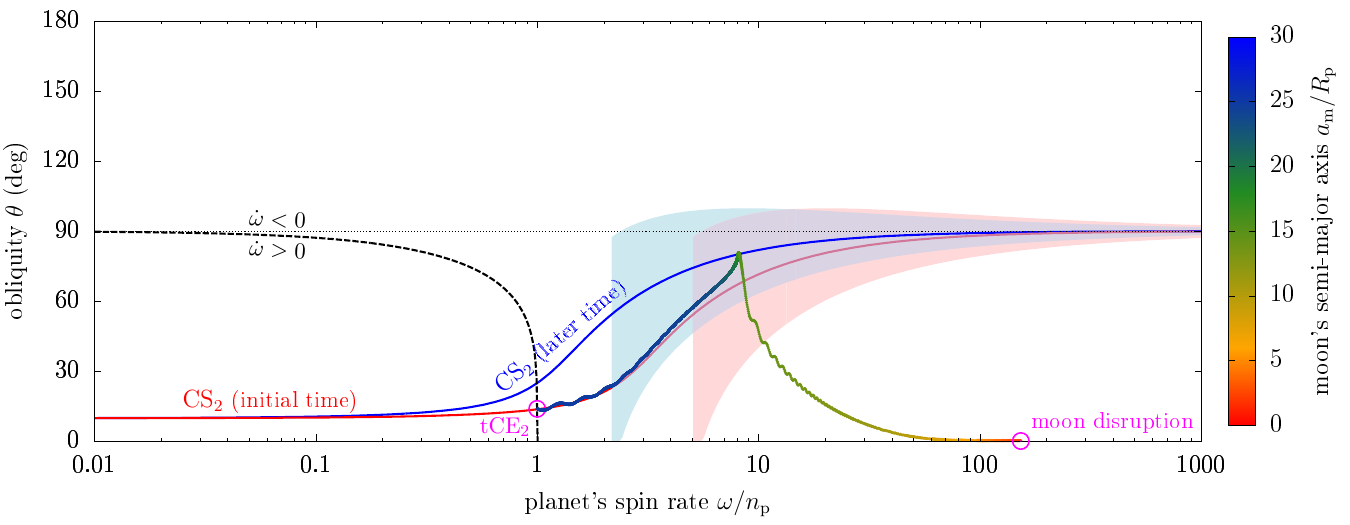}
    \caption{
    Example of evolution towards a temporary high obliquity due to
    inspiral of a moon with mass ratio $m_{\rm m} / m_{\rm p} = 8 \times 10^{-3}$.
    Symbols have the same meaning as in previous figures.
    The system is initialised at $\mathrm{tCE}_2$ and the evolution after the
    moon's disruption is not shown.
    In red, we show the resonance at initial time (for
    $a_\mathrm{m}=25$~$R_\mathrm{p}$); in blue, we show the resonance when
    $\mathrm{CS}_2$ is broken ($a_\mathrm{m}\approx 17$~$R_\mathrm{p}$).
    }\label{fig:bigmoon25}
\end{figure*}

This peculiar effect of massive moons prompts us to investigate whether such moons are able to not only produce long-lived supersynchronous spin states, but also high obliquities. As discussed in Sect.~\ref{sec:theory}, lunar migration can change the
planet's spin-orbit coupling frequency, and introduce new possibilities for resonant
interactions.
In Fig.~\ref{fig:bigmoon25}, we illustrate an example of a planet's spin
evolution under the influence of a massive moon with $m_\mathrm{m}/m_\mathrm{p}=8\times 10^{-3}$.
We adopt the same parameters as for Figs.~\ref{fig:tCE2break_retromoon}-\ref{fig:tCE1break} except that the perturbing planet is located closer in
(at $3$~au), which moves the spin-orbit resonance to the right of the figure.
It can be seen that the initial location of $\mathrm{tCE}_2$ is at a relatively low
obliquity (red curve).
However, as the moon gradually inspirals, two effects occur: (i) $a_\mathrm{m}$
decreases, and (ii) $r_\mathrm{M}$ increases as the planet is spun up.
Both of these effects shift the spin-orbit resonance towards lower values of
$\omega$ (blue curve). Therefore, as the planet adiabatically follows the location of Cassini state~2, its obliquity increases due both to the
moon's migration and to the planet's spin up.
When the moon-induced tidal dissipation becomes too strong, however, the resonance is
broken. The planet then rapidly spins up, and its obliquity damps to zero by the time the
moon is disrupted below the Roche limit.

\begin{figure}
    \centering
    \includegraphics[width=\columnwidth]{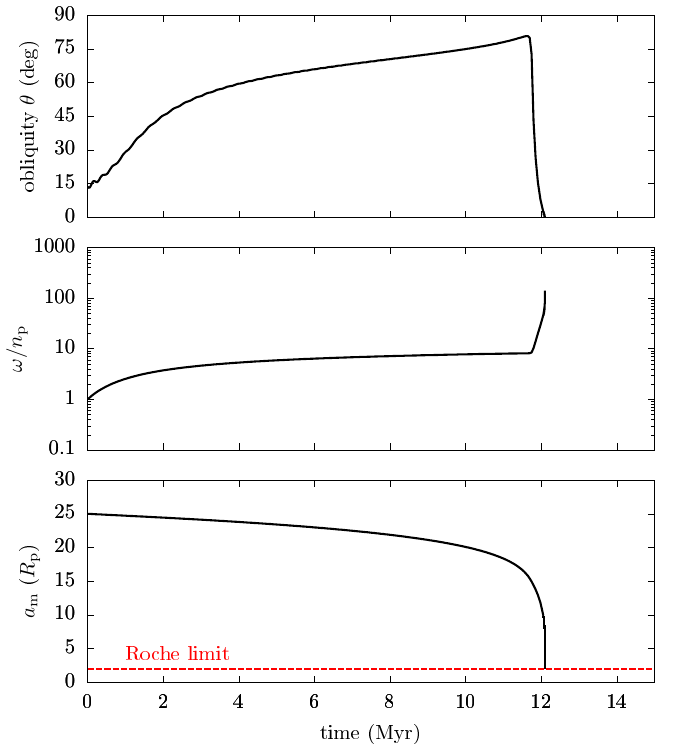}
    \caption{
    Time evolution of the three quantities shown in Fig.~\ref{fig:bigmoon25}:
    the planet's obliquity $\theta$, the planet's spin rate $\omega$, and the
    moon's semi-major axis $a_\mathrm{m}$.
    }\label{fig:bigmoon25_timeevol}
\end{figure}

Even though the obliquity increase discussed here is only temporary, it can
still persist for millions of years.
Figure~\ref{fig:bigmoon25_timeevol} shows the evolution of the planet's
obliquity and spin rate over time.
We see that $\omega$ reaches a long-lived steady state: the
pseudo-synchronization with the moon (see Eq.~\ref{eq:omegapseudo}), for which $\omega/n_\mathrm{m}\approx
2\cos i_\mathrm{m}/(1 + \cos^2i_\mathrm{m})$, where $i_\mathrm{m}$ is the
inclination of the moon's orbit with respect to the planet's equator (where
$i_\mathrm{m}\approx i_{\mathrm{m}1}$ is large, see Eq.~\ref{eq:i_p1}).

The reason for the somewhat short lifetime of the oblique state can be seen in
the bottom panel of Fig.~\ref{fig:bigmoon25_timeevol}: $a_{\rm m}$ continues
decreasing, rather than reaching a tidally-locked equilibrium.
This is because $i_{\rm m}$ generally remains large even after the planet
spins up, since the moon remains outside the planet's Laplace radius:
\begin{align}
    \frac{a_{\rm m, new}}{r_{\rm M}}
        \approx{}&
            25
            \p{\frac{a_{\rm p}}{0.4\;\mathrm{AU}}}
            \p{\frac{R_{\rm p}}{2R_{\oplus}}}^{-1}\nonumber\\
        &\times \p{\frac{m_{\rm p}}{10M_{\oplus}}}^{1/3}
            \p{\frac{m_{\star}}{M_{\odot}}}^{-1/3}
            \p{\frac{m_{\rm m} / m_{\rm p}}{\scinot{5}{-3}}}^{16/45}
        .
\end{align}
As such, the moon remains on a significantly inclined orbit, far from
the tidal equilibrium of the planet-moon system, and the ongoing lunar tide in
the planet results in lunar inspiral after several millions of years.

\section{Conclusion}\label{sec:conclusion}

\subsection{Summary and Discussion}

Moons affect the spin-axis dynamics of their host planets. Even the moons of the Solar System giant planets, which only have moon-to-planet mass ratios of the order of $10^{-4}$ or less, contribute very significantly to the spin-axis precession rate of their planets.
As the orbits of moons expand or contract due to tidal dissipation, moons can have a decisive effect on the planet's spin-axis dynamics, and in particular increase the planet's obliquity.
This intricate dynamical system can lead to complex outcomes, as the ejection or disruption of the moon.
So far, this process has been studied for long-period planets, which are not affected by star-planet tidal dissipation (see e.g. \citealp{Saillenfest-etal_2021,Saillenfest-etal_2023}).

The spin-axis dynamics of short-period planets is largely influenced by star-planet tidal dissipation, which tends to synchronize the planet's spin rate to its orbital frequency and damp its obliquity. Yet, high-obliquity states do exist.
Tidal dissipation creates new kinds of equilibria \citep{su2021dynamics,yuan2024}; it can either have a destabilising or stabilising influence \citep{Su-Lai_2022}. So far, the spin dynamics of short-period planets has been limited to planets without moons, even though moons can readily been formed around short-period planets.

Here, we have investigated the coupled dynamics of warm planets (with periods of about $10$ to $200$ days) and their moons. We have focused both on the spin dynamics of the planet and the fate of its moon. Depending on the hierarchy of timescales between the star-planet tidal dissipation $\tau_\mathrm{p}$ (which synchronises the planet's spin rate and damps its obliquity) and the moon-planet tidal dissipation $\tau_\mathrm{m}$ (which produces the orbital migration of the moon), a variety of possible evolutions can occur.

First, for moderately short-period planets, for which $\tau_\mathrm{p}\gtrsim\tau_\mathrm{m}$, the star-planet tidal dissipation is noticeable but it unfolds on a longer timescale than the moon's migration. In that case, our results show that the planet-moon coupled dynamics described in previous works (see e.g. \citealp{saillenfest2021future,Dbouk-Wisdom_2023}) still applies, but it is accelerated: see Sect.~\ref{sec:longlived} and Fig.~\ref{fig:HIP41378f}. In other words, the migration of the moon is still able to trigger a secular spin-orbit resonance and increase the planet's obliquity, but the migration range needed for the moon is reduced. Consequently, as compared to long-period planets for which $\tau_\mathrm{p}\gg\tau_\mathrm{m}$, a larger proportion of moderately short-period planets should have been tilted and possibly have lost their moons through the final unstable phase (see \citealp{Saillenfest-etal_2022,Saillenfest-etal_2023}). The resulting planets have large obliquities $\theta\gtrsim 80^\circ$ that persist during the lifetime of their planetary systems, because $\tau_\mathrm{p}$ is not short enough to damp the obliquity over billions of years. These findings enlarge the parameter space where moon-driven obliquity excitations are possible, together with the potential formation of very inclined rings (see \citealp{Saillenfest-etal_2023}).

Second, we show that lunar orbital destabilization occurs more readily than predicted by previous works \citep[e.g.][]{barnes2002_moonstab, tremaine2009satellite, saillenfest2021future}.
In particular, the dynamically unstable $\mathrm{E}_1$ region (which occurs for moons with $a_{\rm m} \simeq r_{\rm M}$ orbiting planets with obliquities $\simeq 90^\circ$, \citealp{tremaine2009satellite}) can be reached with the star-planet tidal dissipation alone, even if the moon's migration is negligible.
This is due to a combination of several factors:
First, the Laplace radius, which sets the moon's dynamical regimes, scales as $\omega^{2/5}$, where $\omega$ is the planet's spin rate, so a decrease in $\omega$ is equivalent to an effective outward moon migration.
Second, in the process of damping a planet's obliquity, the star-planet tidal dissipation initially makes the obliquity converge towards $90^\circ$ (see Eq.~\ref{eq:dthetatides}), where the $\mathrm{E}_1$ region lies.
This property implies that short-period planets, for which $\tau_\mathrm{p}\ll\tau_\mathrm{m}$, have a natural tendency to loose their moons by forcing them to cross the $\mathrm{E}_1$ region.
This significantly increases the parameter space where moons are lost compared to existing results \citep{barnes2002_moonstab}:
Fig.~\ref{fig:barnes_comp} summarizes the analytical boundaries between the different regions of the parameter space where each of moon loss dominates.

Third, for moons that survive the planet's star-driven tidal despinning stage of evolution, interesting dynamical effects can then occur.
As $\tau_\mathrm{p}\ll\tau_\mathrm{m}$ for short-period planets, the stellar tides first make the planet converge to one of the two tidal Cassini equilibria, $\mathrm{tCE}_1$ or $\mathrm{tCE}_2$, as they would do for a moon-less planet in a multiplanetary system (see \citealp{su2021dynamics}).
The resulting planet has a slow rotation, close to spin-orbit synchronization, and is in a stable state, either with a low obliquity ($\mathrm{tCE}_1$) or high obliquity ($\mathrm{tCE}_2$).
Yet, the moon slowly migrates inwards due to the planet-moon tidal dissipation with timescale $\tau_\mathrm{m}$.
This stage can last several hundreds of million years, depending on the parameters of the moon, until the moon enters in a phase of runaway inward migration.
At this point, the moon's orbital angular momentum dominates the planet's spin angular momentum (even for small moons with masses $m_\mathrm{m}\lesssim 10^{-4} m_\mathrm{p}$) because the planet spins very slowly.
As a consequence, the runaway moon migration ejects the planet from its tidal Cassini equilibrium and temporarily aligns the planet's spin angular momentum with the moon's orbital angular momentum.
The runaway moon migration ends when the moon is disrupted below the Roche limit, after which the star-planet tidal dissipation makes the planet converge again towards $\mathrm{tCE}_1$ or $\mathrm{tCE}_2$.
This kind of evolution shows that exomoons, even when they no longer exist today, strongly affect the dynamical history of exoplanets.
The migration of moons can eject planets out of spin-orbit equilibria and make them converge to a different equilibrium.
We conclude that the demographics of exomoons represent a crucial uncertainty in theoretical predictions of the present-day obliquities of warm exoplanets.

In addition to small moons, with mass ratios comparable to those of moons around the Solar system giant planets ($m_\mathrm{m}\approx 10^{-4} m_\mathrm{p}$), we have also considered the case of big moons, with mass ratios comparable to that of the Earth's Moon ($m_\mathrm{m}\approx 10^{-2} m_\mathrm{p}$).
For such big moons, the moon-planet tidal dissipation is strong enough to spin the planet all the way up to synchronization with the moon's orbit.
This process is yet another pathway to a high planetary obliquity through a capture in secular spin-orbit resonance.
Contrary to previous scenarios (see e.g. \citealp{Saillenfest-etal_2021}) the obliquity increase here is primarily due to variations in the planet's spin rate and oblateness, and not to the moon's actual migration.
Such high obliquity states are transient, because the moon still slowly migrates inwards towards the Roche limit, but they can be relatively long-lived, of the order of tens of millions of years.

\subsection{Caveats and Future Work}

Throughout this paper, we have assumed that the lunar orbital evolution is driven by the standard constant time lag prescription, and the stellar and lunar tides are modelled self-consistently according to the same prescription.
More complex rheologies and tidal models may result in different evolutionary behaviors compared to those shown in Sections~\ref{sec:e1_instab} and~\ref{sec:despun} (in particular in Figs.~\ref{fig:Kepler79d}--\ref{fig:bigmoon25_timeevol}).
Most importantly, the tidal frequencies of the lunar and planetary orbits differ significantly, suggesting that they may have significantly different dissipation rates when more physically motivated (frequency-dependent) tidal models are adopted.
The impact of adopting such an approach is not immediately clear, and future work will be required to understand whether qualitatively new behaviors are found.
However, we note that some of the other key dynamical evolutionary features necessary for our work are qualitatively present under more complex rheologies, specifically (i) obliquity excitation at rapid spins and obliquity damping at near-synchronous ones \citep{Boue-etal_2016, valente2022tidal} as well as (ii) the presence of high-obliquity tidally-stable equilibria \citep{Boue_2020}.

Additionally, a resonance lock between the lunar orbit and planet's oscillation modes may result in accelerated orbital evolution of the moon, an effect tentatively supported by Solar System observations \citep{fuller2016resonance,
lainey2020resonance}.
Such a mechanism is particularly feasible in our scenario if the moon experiences a resonance lock with an inertial mode of the planet: since inertial mode frequencies scale with the planet's rotation frequency, the planet's spindown would cause the inertial mode frequencies to decrease, raising the lunar semi-major axis \citep{fuller2016resonance}.
While the role of resonance locks in stellar and/or lunar tides is still not fully well-characterized, more realistic tidal models including resonance locks may change the hierarchy of timescales between $\tau_\mathrm{p}$ and $\tau_\mathrm{m}$. The limits of the different regimes derived in this paper and shown in Fig.~\ref{fig:barnes_comp} should therefore not be interpreted as very precise boundaries, but as indicative guidelines for the different regimes of tidal evolution.

Finally, as a third caveat of our tidal model, we neglect cross tides, which arise due to the torque exerted by the tidal bulge raised on the primary by one perturber on a second perturber, as first pointed out in \citet{goldreich1966_moon} but often omitted without mention in subsequent works on lunar dynamics.
The detailed analysis for why this effect can be neglected are somewhat complex (see Appendix~\ref{app:cross_tides}), but the broad reason for its omission is that cross tides are subdominant to the solar-tide-driven planetary spindown as long as the planetary spindown is not significantly slower than the lunar-tide-driven lunar migration.

We have focused here on fluid planets. For rocky planets, their nonzero rigidity would allow them to support $J_2$ moments in excess of their hydrostatic values. For Earth and Venus-like planets, this contribution places a lower bound on their $J_2$ values of $\sim 4 \times 10^{-6}$ \citep{yoder1995book}.
For an Earth-like planet at $0.4\;\mathrm{au}$ as we used here in our fiducial parameters, this would result in a minimum Laplace radius of
\begin{equation}
    r_{\rm M,\min}^5
        = 2\frac{m_{\rm p}}{m_\star}J_2^{\rm(rigidity)}R_{\rm p}^2a_{\rm p}^3
        = (1.82 R_\oplus)^5.
\end{equation}
Around such planets, our assertion that the Laplace radius lies within the planet for tidally despun planets is no longer accurate.
In such cases, the second crossing of the Laplace radius as the moon tidally inspirals may give rise to additional cases of interesting evolution.
However, the maximum asphericity that can be supported by rigidity scales with the dimensionless effective rigidity \citep{Murray-Dermott_1999, zanazzi2017_triax, yuan2024}
\begin{equation}
    \tilde{\mu}
        = \frac{19 \mu}{2\rho g R_{\rm p}}
        \propto \frac{R_{\rm p}^4}{m_{\rm p}^2},
\end{equation}
where $\mu$ is the shear modulus of the planet, $g = Gm_{\rm p}/R_{\rm p}^2$ is its surface gravity, and $\rho_{\rm p}$ its density.
Thus, for planets more massive than a few $M_\oplus$, the maximum $J_2^{\rm (rigidity)}$ decreases sufficiently that $r_{\rm M, \min} \lesssim R_{\rm p}$, and the conclusions in the our study remain qualitatively accurate.

Throughout this work, we have assumed that the putative close-in exoplanets form in their current locations with moon systems resembling those of Solar System planets.
However, orbital migration and even instability are thought to occur in many exoplanetary systems, ideas dating back to the proposed disk-driven migration of hot Jupiters \citep{lin1996_diskHJs} and still relevant in the context of the formation and breaking of planetary resonant chains \citep{izidoro2017_breakchains, li2025_breakchains}.
The survival of exomoons through such an active phase of planet formation remains an open question.
In this vein, a recent work \citep{bolmont2025_moon} considered the survival of exomoons around close-in giant planets during disk-driven migration (paralleling earlier work by \citealp{trani2020_exomoonshj_HEM} for exomoon survival during high-eccentricity migration formation of hot Jupiters).
Our results are complementary to this existing body of literature, focusing on the long-term survival ($\sim$ Gyr) of exomoons after other obstacles to their existence during the early phases of the planet's lifetime.

In this paper, we have not considered tidal dissipation in the moon's interior and the associated circularization of the moon's orbit. This would have required to track the moon's own spin-axis dynamics. With such a more detailed model, our results would remain qualitatively unaffected because the dynamical evolution timescale of the moon (including the destabilization in the $\mathrm{E}_1$ region) is very fast, of the order of $100$~yr, as compared to the millions of years needed for tidal eccentricity damping (see Sect.~\ref{sec:numcode}). Yet, including dissipation in the moon's interior would be needed to determine the actual ultimate fate of the moon when it becomes unstable, whether it is ejected from the planet or disrupted within the Roche limit. In this last case, dedicated simulations (see e.g.\ \citealp{Hyodo-etal_2017}) can be used to assess how the angular momentum of the moon is removed, either through collisions among the debris, or by transfer to the planet.

Finally, we have only considered here the dynamical effects of a single moon. Multiple major moons, as those of Jupiter, would potentially produce more complex dynamical evolutions, with inner moons migrating faster and dynamically interacting with outer moons. If a given moon is destabilized by passing through the $\mathrm{E}_1$ region, it could also produce a chain reaction among other moons and generate a complicated reorganization of the moon system. These possibilities open the door to many more interesting evolution pathways, and draw us back to the conclusion that the possible existence of exomoons, for which we have no statistics yet, is a confounding factor in obliquity predictions.

\begin{acknowledgements}
      We thank the referee for detailed and insightful comments that improved the quality of this manuscript.
      YS thanks Daniel Fabrycky, Gongjie Li, and Scott Tremaine for useful discussions.
      YS acknowledges support by the Lyman Spitzer Jr.\ Postdoctoral Fellowship at Princeton University and by the Natural Sciences and Engineering Research Council of Canada (NSERC) [funding reference CITA 490888-16].
\end{acknowledgements}

\bibliographystyle{aa} 
\bibliography{bib} 

\appendix

\section{Cross Tides}\label{app:cross_tides}

A rather subtle yet important caveat in our work is the omission of the so-called ``cross tidal'' interactions \citep{goldreich1966_moon, touma1994_ctides, desurgy1997_ctides, mathis2009_crosstide}.
While these interactions are often neglected without explicit consideration, they are known to significantly affect the history of the lunar orbit, as first pointed out by the seminal work of \citet{goldreich1966_moon}.
Here, we present an abbreviated discussion of cross tides, focusing on the key results necessary for justifying their omission.

In the standard tidal theory, the [time-lagged] bulge raised by a perturbing body exerts a torque on that same perturber, driving dissipation.
However, in systems with at least three bodies, the bulges raised on the primary by one perturber also exert torques on the other perturber[s].\
Readers interested in a more complete discussion of this phenomenon are encouraged to refer to \citet{touma1994_ctides}, which contains the most complete expressions for the architecture we consider: they do not average over the lunar orbital precession as do \citet{goldreich1966_moon} and \citet{desurgy1997_ctides}, which is only appropriate for moons beyond the Laplace radius, such as Earth's moon today.

The essential effect of cross tides can be understood simply.
As above, we consider a three-body system consisting of the star, planet, and moon.
In the frame of the planet, the two perturbers are the star, which orbits with
frequency $n_{\rm p}$, and the moon, which orbits with frequency $n_{\rm m}$.
We assume that the system is not near any resonances, so $n_{\rm p}$ and $n_{\rm
m}$ are not commensurate.
In each cross tidal contribution, the bulge raised by one body, the
``perturbing'' body, effects a torque on the other body, the ``torqued'' body.
Then, when averaging over the mean anomalies of both orbits, it is clear that
only the axisymmetric components of the gravitational potentials of both the perturbing and torqued bodies can interact in a way that results in a non-vanishing torque.
While it is not obvious that time-lagged axisymmetric perturbations can lead to
a net torque, this indeed is possible when the two
orbits are misaligned (e.g. see \citealp{lai2012} for another example of this).

By recognizing the independence of the cross tidal torques of the bodies' mean
anomalies, we identify one of their key features: they cannot change the
magnitudes of the orbital angular momenta in the system (notably, they do not
yield lunar migration; see also the explicit torques provided by \citealp{touma1994_ctides}).
Thus, the cross tide results in the planet's spin axis and torqued body's orbit
normal being driven towards alignment at constant total angular momentum and
constant orbital angular momentum magnitude.
A complementary decrease of the planet's spin rate is a consequence of these two
constraints.
The characteristic evolution timescale of the planet's spin rate due to cross
tides can be written \citep{goldreich1966_moon, touma1994_ctides}
\begin{align}
    \frac{\tau_{\rm p, cross-tide}}{\tau_{\rm p}}
        ={}&
            \frac{m_\star}{m_{\rm m}}\p{\frac{a_{\rm m}}{a_{\rm
                p}}}^3\label{eq:tau_crosstide}\\
        ={}&
            3.21
                \p{\frac{m_{\rm m} / m_{\rm p}}{10^{-4}}}^{-1}
                \p{\frac{m_{\rm p}}{10M_{\oplus}}}^{-1}
                \p{\frac{m_\star}{M_{\odot}}}\nonumber\\
            &\times
                \p{\frac{a_{\rm m} / R_{\rm p}}{10}}^{3}
                \p{\frac{R_{\rm p}}{2R_{\oplus}}}^{3}
                \p{\frac{a_{\rm p}}{0.4\;\mathrm{au}}}^{-3},\nonumber
\end{align}
where $\tau_{\rm p}$ as given by Eq.~\eqref{eq:taup} is the despinning timescale
due to the bulge raised by the star torquing the star.
The instantaneous despinning time due to cross tides includes some prefactors,
but they generally result in a longer spin evolution timescale than predicted by
Eq.~\eqref{eq:tau_crosstide}, approaching infinity as the misalignment in the
system goes to zero (where there is no cross-tidal torque).

Towards understanding the effect of cross tides on our results, we turn to a timescale analysis.
Note that
\begin{equation}
    \tau_{\rm p, cross-tide} = \sqrt{\tau_{\rm p} \tau_{\rm p, m}},
\end{equation}
where $\tau_{\rm p, m}$ is the despinning time due to the lunar bulge torquing the moon (Eq.~\ref{eq:tau_pm}).
Since $\tau_{\rm p, m} \sim \tau_{\rm m} (S_{\rm p} / L_{\rm m})$, we see that $\tau_{\rm p, cross-tide} \lesssim \tau_{\rm p}$ requires $\tau_{\rm m} \lesssim \tau_{\rm p} (L_{\rm m} / S_{\rm p}) \ll \tau_{\rm p}$.
In other words, requiring that the cross-tidal despinning timescale be non-negligible also requires that the lunar migration time be significantly shorter (even more so than the angular momentum ratio $L_{\rm m} / S_{\rm p}$) than the planetary spin evolution timescale.
Since the focus of this paper is on regimes where the planetary spin evolution is at least as fast as the lunar migration, we conclude that the cross-tidal evolution is negligible in affecting the planet's spin evolution in our study.
In conjunction with the lack of an effect on the lunar orbit, we justify the omission of cross tides in our study.

\section{Evection and eviction resonances}\label{asec:evec}

Evection and eviction resonances happen when the apsidal or nodal precession frequencies of a moon are commensurable with the mean motion of its host planet on its orbit around the star
\citep{Touma-Wisdom_1998,Vaillant-Correia_2022}. These resonances can substantially affect the eccentricity and inclination of a moon; however, these resonances are not included in secular models that are averaged over the orbital motion of the planet. In order to use a secular code (as we do in all the numerical experiments in this paper), we must assert that these resonances cannot be reached, or cannot appreciably change the dynamics at play.

In this paper, we only consider relatively small moons ($m_\mathrm{m}/m_\mathrm{p}$ of the order of $10^{-4}$) orbiting close to their Laplace equilibrium (see \citealp{tremaine2009satellite}). The close proximity to the Laplace equilibrium is indeed verified at all times in our simulations, except when the system enters the $\mathrm{E}_1$ region, in which case the moon is rapidly destabilized anyway (see Sect.~\ref{sec:e1_instab}). Therefore, outside of the $\mathrm{E}_1$ region, a good approximation to the apsidal and nodal precession frequencies of the moon is given by the frequencies of small oscillations around the Laplace equilibrium. These frequencies have been computed by \cite{tremaine2009satellite} for a massless moon. We call them $\mu_1$ and $\xi_1$, respectively, following Eqs.~(19) and (16) of \cite{saillenfest2021future}.

\begin{figure}
   \includegraphics[width=\columnwidth]{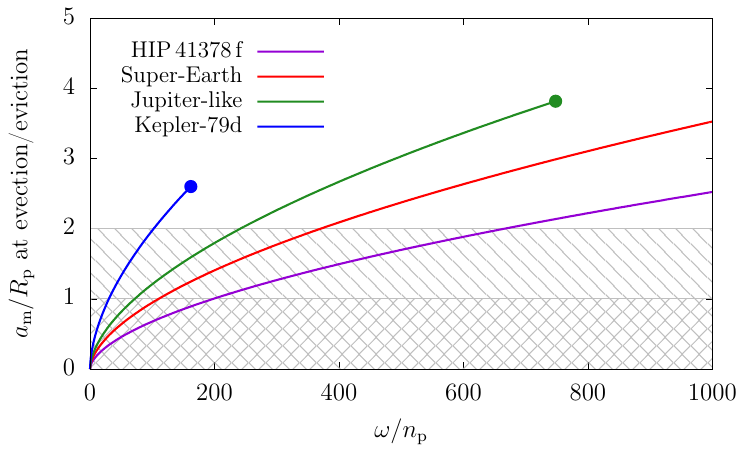}
   \caption{Distance of the moon at the evection (or eviction) resonance as a function of the spin rate of the planet. The four curves cover all different settings considered throughout this paper. On each curve, the dot shows the planet's breakup spin rate. The two hatched zones below $a_\mathrm{m}/R_\mathrm{p}=1$ and $a_\mathrm{m}/R_\mathrm{p}=2$ show the planet's radius and approximate Roche limit, respectively.}
   \label{fig:ev}
\end{figure}

Evection and eviction-like resonances are triggered when the planet's mean motion $n_\mathrm{p}$ is in an integer commensurability with $\mu_1$ or/and $\xi_1$. \cite{Vaillant-Correia_2022} give a list of all possible commensurabilities in the quadrupolar approximation. For a given spin rate $\omega$ of the planet, we can locate where these commensurabilities occur in the plane $(a_\mathrm{m},\theta)$. By doing so, we note that, for the parameters considered in this paper, evection and eviction-like resonances are all located at distances $a_\mathrm{m}$ significantly closer to the planet than the mid-point radius $r_\mathrm{M}$. Consequently, we can simplify the expression of $\mu_1$ and $\xi_1$ to
\begin{equation}
   \mu_1 = -\xi_1 = \frac{3}{2}n_\mathrm{m}J_2\frac{R_\mathrm{p}^2}{a_\mathrm{m}^2}\,,
\end{equation}
(see Eqs.~B.2 and B.5 of \citealp{saillenfest2021future}). By equating $\mu_1$ (or $-\xi_1$) with $n_\mathrm{p}$, we get the nominal location of the evection (or eviction) resonance:
\begin{equation}\label{eq:ev}
   \frac{a_\mathrm{m}}{R_\mathrm{p}} = \left[\frac{k_2^2}{4}\frac{m_\star}{m_\mathrm{p}}\left(\frac{R_\mathrm{p}}{a_\mathrm{p}}\right)^3\left(\frac{\omega}{n_\mathrm{p}}\right)^4\right]^{1/7}\,.
\end{equation}
As a function of the spin rate of the planet, the resulting distance of the moon is shown in Fig.~\ref{fig:ev}. We note that, as the planet spins down due to the star-driven tidal dissipation, the location of the evection resonance rapidly drifts below the planet's own radius. This is because $J_2$ decreases, which slows down the moon's orbital precession. As a result, the evection or eviction resonances can never be reached by the moon.

Figure~\ref{fig:ev} only shows the location of the resonance $n_\mathrm{p}+\xi_1=0$. Among the other resonances listed by \cite{Vaillant-Correia_2022}, some are located farther away from the planet. If we consider the most distant resonance, this adds a factor $3^{2/7}\approx 1.37$ to Eq.~\eqref{eq:ev}. This distance still remains firmly below the distance of the moon at any point during its evolution. Moreover, for moons close to their Laplace equilibrium, with nearly circular orbits, \cite{Vaillant-Correia_2022} note that only the resonances $n_\mathrm{p}+\xi_1=0$ and $2n_\mathrm{p}+\xi_1=0$ can have a substantial effect (and the resonance $2n_\mathrm{p}+\xi_1=0$ is located even closer-in than shown in Fig.~\ref{fig:ev}).

From this discussion, we conclude that evection and eviction-like resonances can safely be ignored throughout this paper. This justifies the use of a secular code averaged over the orbital motion of the planet.

\end{document}